\documentclass[12pt,a4paper]{article}
\usepackage{babel}
\usepackage{graphicx} 
\usepackage{tikz}
\usepackage{multirow}
\usepackage{pdflscape}
\usepackage{amsmath}
\usepackage{hyperref}
\usepackage{caption}
\usepackage{lscape}
\usepackage{tabularx,ragged2e,colortbl}
\usepackage[normalem]{ulem}
\usepackage{array, makecell}
\usepackage{centernot}
\usepackage[left=2.5cm,right=2.5cm,top=2.5cm,bottom=2.5cm]{geometry}
\usepackage{comment}

\usepackage[mathlines, pagewise]{lineno}

\usepackage[numbers, sort&compress]{natbib}
\usepackage{xcolor}
\newcommand{\blue}[1]{\textcolor{blue}{#1}}

\newcommand{\red}[1]{\textcolor{red}{#1}}

\newcommand{\indep}{\perp \!\!\! \perp}
\newcommand{\nindep}{\not\!\perp\!\!\!\perp}

\newcommand{\E}{\mathrm{E}}
\newcommand{\PP}{\mathrm{P}}

\usepackage{newfloat}
\DeclareFloatingEnvironment{Supplementary Figure}

\DeclareFloatingEnvironment{Supplementary Table}

\title{Using Non-Targeted HPV Infections in Studies with Risk Compensation}

\begin{document}

\thispagestyle{empty}

\maketitle

\large
\begin{center}
Lola Etiévant$^{1*}$
\end{center}

\normalsize

\noindent $^{1}$ Univ. Grenoble Alpes, CNRS, LJK, 38000 Grenoble, France.\\

\noindent $^{*}$ Corresponding author

\noindent \href{mailto:lola.etievant@univ-grenoble-alpes.fr}{lola.etievant@univ-grenoble-alpes.fr} \quad \quad \href{https://orcid.org/0000-0001-7562-3550}{ORCID 0000-0001-7562-3550}

\noindent SVH Team, DATA Department, Jean Kuntzmann Laboratory, Bâtiment IMAG, 150 place du Torrent, 38401 Campus Universitaire de Saint-Martin-d'Hères, FRANCE\\

\noindent The final published article and its Supplementary Material are available at the \red{\href{https://www.degruyterbrill.com/document/doi/10.1515/ijb-2025-0087/html}{lnternational Journal of Biostatistics website}}.


\newpage

\thispagestyle{empty}

\section*{Abstract}

Studies of HPV vaccine efficacy usually record infections with vaccine targeted and non-targeted strains. Contrary to blinded randomized controlled trials, confounding bias can be a threat and risk compensation may occur in observational studies. Etievant et al. (Biometrics, 2023) proposed to use cervical infections with non-targeted HPV strains to remove or reduce confounding bias in estimates of vaccine efficacy on targeted strains. However, they assumed that vaccinated women could not change their behavior after vaccination. This work investigates if the quantity estimated in practice with the method of Etievant et al. has a clear causal meaning under a more plausible setting where unmeasured sexual behavior acts as both a confounder and a mediator. Under certain assumptions, using non-targeted HPV infections can remove both confounding bias and the portion of the vaccine effect on the targeted HPV strains that is mediated through the change of behavior. In that case, the estimated quantity has a clear causal interpretation as it represents the direct immunological effect of the vaccine. Infections with non-targeted HPV strains could also be used to isolate the indirect behavioral effect of the vaccine in unblinded randomized controlled trial.\\ 


\noindent \textbf{Keywords:} HPV vaccine; mediation analysis; non-targeted HPV strains; risk compensation; observational studies; randomized controlled trials\\

\newpage

\setcounter{tocdepth}{2}
\setcounter{page}{1}

\section{Introduction}\label{sec:intro}

There are more than 100 types of human papillomavirus (HPV), some of which can cause cancer, notably cervical cancer. Several HPV vaccines have been developed; they target the two most carcinogenic strains, HPV 16 and 18, and up to seven other HPV strains (e.g., HPV 6 and 11). Both randomized controlled trials (RCTs) and observational studies are conducted to study the effect of HPV vaccine against infection by the targeted HPV strains \citep{escuddo2022, basu2021observational}. 

In a RCT, the vaccine is randomly assigned to, say, half of the participants, who constitute the treatment group, while the other half receives a placebo and constitutes the control group. Trial participants are then followed and the ratio (or difference) of incident cervical targeted HPV infections in the two groups is interpreted as the causal effect of the vaccine. In particular, the randomized attribution of the vaccine preserves the estimation from confounding bias; thanks to randomization, the two groups are ``exchangeable'' in the sense that they should have the same characteristics before vaccination. 

But notably for ethical reasons, epidemiologists sometimes only have access to observa-tional data. Appropriate estimation of the causal effect of the vaccine on targeted HPV infections then requires adjustment for factors impacting both vaccination and the risk of infection (i.e., confounders of the exposure-outcome relationship). Alas, one can expect vaccine effect estimates to be biased as key confounders such as sexual behavior are usually mismeasured or unmeasured \citep{fewell2007confbias, diclemente2016sexualbehavior}. Fortunately, cervical infections with the non-targeted HPV strains share the same transmission route as the targeted strains \citep{mendez2005coinfection, chaturvedi2011coinfection} and are often recorded in HPV vaccine studies. Making the common assumption that non-targeted strains that are phylogenetically unrelated to targeted strains are unaffected by vaccination, Etievant et al. \cite{etievantsampson2023} proposed to use the non-targeted HPV infections as a negative control outcome (NCO) \citep{lipsitch2010NCO} to remove or substantially reduce confounding bias from the estimated vaccine effect on the targeted infections. 

Another advantage of RCTs of vaccine efficacy is that they are usually blinded, preventing participants from knowing if the vaccine or the placebo was administered, and therefore precluding behavior changes in response to a perceived decrease in risk, a phenomenon known as risk compensation \citep{wilde1998riskhomeostasis, hedlund2000riskcompensation}. Etievant et al. \cite{etievantsampson2023} assumed that in observational settings, factors impacting the risk of HPV infections such as unmeasured sexual behavior only acted as confounders. However, while it is plausible that depending on their sexual behavior women may be more or less likely to get the HPV vaccine, one could also imagine that women could possibly engage in riskier sexual behaviors after vaccination as they think they are somehow protected against HPV infections \citep{brewer2007riskcompensationvaccine}. Then, under such a setting, the vaccine would have a direct (i.e., immunological) effect on the risk of targeted HPV infections, and an indirect effect (i.e., through the change of sexual behavior). And because of the common mode of transmission of cervical HPV infections, the vaccine would also have a causal effect on the non-targeted HPV infections. This would consequently violate the key assumption in \cite{etievantsampson2023}.

Although empirical evidence suggests limited change in sexual behavior in women after HPV vaccination, it is difficult to properly assess risk compensation \citep{mullins2012HPVbehavior, underhill2013HIVprevention, donken2018HPVriskcompensation, amboree2023HPVbehavior}. In this work, such a plausible observational setting is considered, with unmeasured sexual behavior acting as both a confounder and a mediator in the relationship between the vaccine and the targeted HPV infections. The objective is to investigate if the quantity estimated in practice with the method proposed by Etievant et al. \cite{etievantsampson2023} has a clear causal meaning under this setting too. The notation and a formal directed acyclic graph are first presented in Section \ref{sec:method}. Section \ref{sec:method} then examines how the quantity estimated in practice is related to the average total effect and to the natural direct and indirect effects. The approach is evaluated in a simulation in Section \ref{sec:simul}. Concluding remarks are presented in Section \ref{sec:discussion}. A discussion on the potential use of non-targeted HPV infections in randomized settings to assess risk compensation is presented in the Supplementary Material, along with technical derivations and details.

\section{Method}\label{sec:method}

\subsection{General setting and notation}\label{subsec:notation}

Similar notation as in \cite{etievantsampson2023} are used. To simplify, a setting without measured confounders is first considered. Let $n$ denote the number of women in the observational HPV vaccine study. Then let $T_i$ and $Y_{1,i}$ respectively denote the binary indicator of vaccination and the binary indicator of cervical infection by a targeted HPV strain, say, 16, for subject $i \in \{1, \dots , n \}$. Assume that infections with $N_{NT}$ HPV strains that are non-targeted by the vaccine and phylogenetically unrelated to the targeted strains were recorded. Let $Y_{2,i}^{(j)}$ denote the binary indicator of infection by the $j^{\text{th}}$ non-targeted strain for subject $i$, $j \in \{1, \dots , N_{NT}\}$, and $Y_{2,i} = \sum_{j=1}^{N_{NT}} Y_{2,i}^{(j)}$ be the total number of non-targeted HPV infections for subject $i$. Finally, let $\boldsymbol{A}_i$ denote the values of the set of unobserved confounders for subject $i$, and $\boldsymbol{\tilde A}_i$ be the values of the set of same variables \underline{after vaccination}. Typically, $\boldsymbol{A}$ and $\boldsymbol{\tilde A}$ are related to sexual behavior. A more general setting with a set of observed confounders, $\boldsymbol{W}$, is considered in Section \ref{subsec:conf}. Assume the relationships among the variables are as in the directed acyclic graph in Figure \ref{fig:DAGmediation} \textbf{(A)}, with the following temporal ordering of the variables: $\boldsymbol{A}, T, \boldsymbol{\tilde{A}}, Y_1$ and $Y_2$. More precisely, assume that:
\begin{eqnarray}
    \E(Y_{1,i} \mid T_i, \boldsymbol{\tilde{A}}_i) &=& g_1(\boldsymbol{\tilde{A}}_i) \exp(\alpha_1 + \beta_1 T_i),\label{eq:equation1}\\
    \E(Y_{2,i} \mid T_i, \boldsymbol{\tilde{A}}_i) &=& g_2(\boldsymbol{\tilde{A}}_i) \exp(\alpha_2).
    \label{eq:equation2}
\end{eqnarray}
Equations (\ref{eq:equation1}) and (\ref{eq:equation2}) are similar to that in \cite{etievantsampson2023} (when $\boldsymbol{W} = \emptyset$), but a few remarks follow. Importantly, $T$ is assumed not to affect $Y_{2}$ directly, but only indirectly, through $\boldsymbol{\tilde{A}}$. On the other hand, $T$ affects $Y_{1}$ both directly and indirectly. Again, as in \cite{etievantsampson2023}, if a certain behavior increases the risk of an HPV 16 infection, one can reasonably assume it is likely to increase the risk of non-targeted infections proportionally, as the targeted and non-targeted strains share the same transmission route. In that case, one would assume (Assumption A$_1$) $g_2 = k \times g_1$ with $k>0$. When $k \neq 1$, only the shape, but not the magnitude, of the effect of $\boldsymbol{\tilde{A}}$ is the same for targeted and non-targeted infections. Additional remarks on the choice of models are made in Supplementary Material \ref{appendix:models}.

\subsection{Causal effects of potential interest}\label{subsec:causaleffects}

Compared to the setting in \cite{etievantsampson2023}, $T$ also has an effect on $Y_1$ through $\boldsymbol{\tilde{A}}$; see Figure \ref{fig:DAGmediation} \textbf{(A)}. In other words, $\boldsymbol{\tilde{A}}$ is a mediator in the relationship between $T$ and $Y_1$. The causal effect of $T$ on $Y_1$ can thus be decomposed into the portion that is mediated by $\boldsymbol{\tilde{A}}$, the ``indirect'' effect, and the portion that is not mediated by $\boldsymbol{\tilde{A}}$, the ``direct'' effect. Just as for the total causal effect of the vaccine, these effects can be defined from counterfactual variables \citep{RobinsGreenland1992, Pearl2001}; see below.

Additional notation to formally define these causal effects is now introduced. Let $Y_{1,i}^{1}$ denote what would have been observed for the $Y_1$ variable if subject $i$ would have been vaccinated, and $Y_{1,i}^{0}$ denote what would have been observed if subject $i$ would have not received the vaccine, $i \in \{1, \dots , n \}$. Similarly, $\boldsymbol{\tilde{A}}_i^{t}$ is the counterfactual value for $\boldsymbol{\tilde{A}}$ if the treatment for subject $i$ would have been set to $t$, $t \in \{0,1\}$. Finally, $Y_{1,i}^{t, \boldsymbol{\tilde{A}}_i^{t^*}}$ is the value for $Y_1$ if $T_i$ would have been set to $t$, and if $\boldsymbol{\tilde{A}}_i$ would have been set to the value it would have taken in the counterfactual world where $T_i$ would have been set to $t^*$, $t, t^* \in \{0,1\}$. 

Because it is generally not possible to identify individual causal effects, population causal effects are considered. First, the average total effect of the vaccine on the targeted HPV infections, on the ratio scale, is defined as $ATE(Y_1) = \frac{{\E} \left( Y_1^{1} \right)}{\E \left(Y_1^{0} \right)}$. It compares the risk of getting a targeted HPV infection, had all the women been vaccinated, to the risk of getting a targeted HPV infection, had all the women not received the vaccine. As a remark, when using the ratio scale, the \textit{vaccine efficacy} is usually defined as $1 - \frac{{\E} \left( Y_1^{1} \right)}{\E \left(Y_1^{0} \right)}$; one could also consider the log-ratio effect, $\log \left\{\frac{{\E} \left( Y_1^{1} \right)}{\E \left(Y_1^{0} \right)} \right\}$. Observe that
\begin{eqnarray*}
  ATE(Y_1)  &=& \frac{\E \left( Y_1^{1,\boldsymbol{\tilde{A}}^{1}} \right)}{\E \left(Y_1^{0, \boldsymbol{\tilde{A}}^{0}} \right)},  \hspace{3cm}  \\
 \hspace{3cm} &=& \blue{\frac{\E \left( Y_1^{1, \boldsymbol{\tilde{A}}^{1}}\right)}{\E \left(Y_1^{1, \boldsymbol{\tilde{A}}^{0}} \right)} }\times \red{\frac{\E \left( Y_1^{1, \boldsymbol{\tilde{A}}^{0}}\right)}{\E \left(Y_1^{0, \boldsymbol{\tilde{A}}^{0}} \right)}}, \\
    &:=& \blue{NIE(Y_1, T = 1)} \times \red{NDE(Y_1, T = 0)}.
\end{eqnarray*}
The second contrast on the right-hand side of the above equation defines the natural direct effect and quantifies the causal effect of $T$ on $Y_1$ not through $\boldsymbol{\tilde{A}}$, i.e., the effect of $T$ on $Y_1$ if for each woman $\boldsymbol{\tilde{A}}$ would have been set to $\boldsymbol{\tilde{A}}^{0}$. The first contrast defines the natural indirect effect (under $T = 1$) and quantifies the remaining portion of the effect, i.e., the effect of $T$ on $Y_1$ only along the path $T \rightarrow \boldsymbol{\tilde{A}} \rightarrow Y_1$. In this work, the direct effect constitutes the immunological effect of the vaccine, while the indirect effect is due to the change of behavior following vaccination. Similarly, the log-ratio effect can be decomposed as $\blue{ \log \left\{ \frac{\E \left( Y_1^{1, \boldsymbol{\tilde{A}}^{1}}\right)}{\E \left(Y_1^{1, \boldsymbol{\tilde{A}}^{0}} \right)} \right\} } + \red{ \log \left\{\frac{\E \left( Y_1^{1, \boldsymbol{\tilde{A}}^{0}}\right)}{\E \left(Y_1^{0, \boldsymbol{\tilde{A}}^{0}} \right)} \right\} }$. 

Several conditions are needed to identify $\E \left( Y_1^{t} \right)$ and $E \left(Y_{1}^{t,\boldsymbol{\tilde{A}}^{t^*}} \right)$, $t, t^* \in \{0,1\}$, and thus $ATE(Y_1)$, $NIE(Y_1,T=1)$, and $NDE(Y_1, T = 0)$. Under the setting of Figure \ref{fig:DAGmediation} \textbf{(A)} and given data on $\{\boldsymbol{{A}}, T,\boldsymbol{\tilde{A}}, Y_1 \}$, these causal effects could be estimated in practice; see Supplementary Material \ref{appendix:identifiability}. Specifically, under Figure \ref{fig:DAGmediation} \textbf{(A)} and using Equation (\ref{eq:equation1}),
\begin{eqnarray}
    ATE(Y_1) &=& \exp(\beta_1) \times \frac{\E_{\boldsymbol{A}}\big[\E\big\{ g_1(\boldsymbol{\tilde{A}})\mid \boldsymbol{A}, T=1\big\} \big]}{\E_{\boldsymbol{A}}\big[\E \big\{ g_1(\boldsymbol{\tilde{A}})\mid \boldsymbol{A}, T=0\big\} \big]},\label{eq:ATE}
\end{eqnarray}
\begin{eqnarray*}
    NIE(Y_1,T=1) = \frac{\E_{\boldsymbol{A}}\big[\E\big\{ g_1(\boldsymbol{\tilde{A}})\mid \boldsymbol{A}, T=1\big\} \big]}{\E_{\boldsymbol{A}}\big[\E \big\{ g_1(\boldsymbol{\tilde{A}})\mid \boldsymbol{A}, T=0\big\} \big]},
\end{eqnarray*}
and
\begin{eqnarray}
    NDE(Y_1, T = 0) = \exp(\beta_1).\label{eq:NDE}
\end{eqnarray}

\subsection{Causal interpretation of the contrast estimated in practice}\label{sec:contrast}

In practice neither $\boldsymbol{{A}}$ nor $\boldsymbol{\tilde{A}}$ are measured. Given data on $\{T, Y_1, Y_2\}$, the contrast one would probably estimate in an attempt to quantify the effect of the vaccine on the targeted HPV strain is $\frac{\E(Y_{1} \mid T = 1)}{\E(Y_{1} \mid T = 0)}$. But using Equation (\ref{eq:equation1}), it follows that
\begin{eqnarray*}
\E(Y_{1} \mid T ) = \E_{\boldsymbol{\tilde{A}}} \big\{ \E\big( Y_1 \mid \boldsymbol{{\tilde{A}}}, T \big)\mid T\big\} = \exp(\alpha_1+ \beta_1 T) \times \E \big\{ g_1(\boldsymbol{\tilde{A}})\mid T\big\},
\end{eqnarray*}
and thus
\begin{eqnarray*}
\frac{\E(Y_{1} \mid T = 1)}{\E(Y_{1} \mid T = 0)} =\exp(\beta_1) \times \frac{\E\big\{ g_1(\boldsymbol{\tilde{A}})\mid T=1\big\}}{\E \big\{ g_1(\boldsymbol{\tilde{A}})\mid T=0\big\}}.
\end{eqnarray*}
Clearly, this contrast does not coincide with $ATE(Y_1)$, $NIE(Y_1,T=1)$, or $NDE(Y_1, T = 0)$. Recall that it differs from $ATE(Y_1)$ because of confounding bias, as $\boldsymbol{{A}}$ is a confounder for the $T-Y_1$ relationship \citep{etievantsampson2023}. As a result, this contrast does not have a clear causal meaning.

\subsection{Using infections with non-targeted HPV strains}\label{subsec:nontargeted}

A common assumption is that the HPV vaccine does not have an effect on the non-targeted strains that are phylogenetically unrelated to the targeted strains \citep{sasieni2022HPVnco, etievantsampson2023}. In the present setting, it is assumed that vaccination does not have an immunological effect on them, but can have an indirect effect through the change of sexual behavior. Therefore, under the setting of Figure \ref{fig:DAGmediation} \textbf{(A)}, $\boldsymbol{\tilde{A}}$ is also a mediator in the relationship between $T$ and $Y_2$, and $T$ thus has a causal effect on $Y_2$. Then, $Y_2$ is not a valid NCO for $T$ and $Y_1$ \citep{lipsitch2010NCO, etievantsampson2023}. Note, the causal effect of $T$ on $Y_2$ is fully mediated by $\boldsymbol{\tilde{A}}$. 

However, using Equation (\ref{eq:equation2}) one knows that, for any $t \in \{0,1\}$,
\begin{eqnarray*}
E(Y_{2} \mid T ) = \E_{\boldsymbol{\tilde{A}}} \big\{ \E\big( Y_2 \mid \boldsymbol{{\tilde{A}}}, T \big)\mid T\big\} = \exp(\alpha_2) \times \E \big\{ g_2(\boldsymbol{\tilde{A}})\mid T\big\},\label{eq:contrastY2}
\end{eqnarray*}
so that $\frac{\E(Y_{2} \mid T = 1)}{\E(Y_{2} \mid T = 0)} = \frac{\E \big\{ g_2(\boldsymbol{\tilde{A}})\mid T=1\big\}}{\E \big\{ g_2(\boldsymbol{\tilde{A}})\mid T=0\big\}}$. Then, under Assumption A$_1$ (given below and motivated in Section \ref{subsec:notation}), $\frac{\E(Y_{1} \mid T = 1)}{\E(Y_{1} \mid T = 0)} \times \frac{\E(Y_{2} \mid T = 0)}{\E(Y_{2} \mid T = 1)} = \exp(\beta_1)$ and thus coincides with $NDE(Y_1, T = 0)$. In other words, under the setting of Figure \ref{fig:DAGmediation} \textbf{(A)}, using the contrast estimated from the non-targeted infections as in method Joint-NC \citep{etievantsampson2023} allows to ``correct'' that of the targeted infections so that it has clear causal meaning. More precisely, in this setting it removes the confounding bias due to differences in sexual behavior pre-vaccination and the indirect effect of the vaccine through the change of behavior after vaccination. In this way, the quantity estimated in practice amounts only for the immunological effect of the vaccine.  
$$\text{A}_1: \quad \text{ There exists }k > 0 \text{ such that } g_2(\boldsymbol{\tilde{a}}) = k \times g_1(\boldsymbol{\tilde{a}}) \text{ for all possible values } \boldsymbol{\tilde{a}} \text{ of } \boldsymbol{\tilde{A}}.$$
As also proposed by Etievant et al. \cite{etievantsampson2023}, a variance estimate can be obtained from the sandwich formula based on the joint estimating equation for $Y_1$ and $Y_2$; see Supplementary Material \ref{appendix:estimationinference}.

\subsection{Setting with measured confounders}\label{subsec:conf}

A more general setting is now considered. The same notation as in Section \ref{subsec:notation} is used, but $\boldsymbol{W}_i$ denotes the values of the set of observed confounders for subject $i \in \{1, \dots , n \}$. The relationships among the variables are as in Figure \ref{fig:DAGmediation} \textbf{(B)}, with the following temporal ordering of the variables: $\boldsymbol{W}, \boldsymbol{A}, T, \boldsymbol{\tilde{A}}, Y_1$ and $Y_2$. More precisely, and similarly as in \cite{etievantsampson2023}, assume that:
\begin{eqnarray}
    \E(Y_{1,i} \mid \boldsymbol{W}_i, T_i, \boldsymbol{\tilde{A}}_i) &=& g_1(\boldsymbol{\tilde{A}}_i) \exp\{\alpha_1 + \beta_1 T_i + s_1(\boldsymbol{W}_i)\},
    \label{eq:equation1w}\\
    \E(Y_{2,i} \mid \boldsymbol{W}_i, T_i, \boldsymbol{\tilde{A}}_i) &=& g_2(\boldsymbol{\tilde{A}}_i) \exp\{\alpha_2 + s_2(\boldsymbol{W}_i)\}.
    \label{eq:equation2w}
\end{eqnarray}

Under the setting of Figure \ref{fig:DAGmediation} \textbf{(B)} and given data on $\{\boldsymbol{W}, \boldsymbol{{A}}, T,\boldsymbol{\tilde{A}}, Y_1 \}$, the average total effect, natural direct effect and natural indirect effect could be estimated; see Supplementary Material \ref{appendix:identifiabilityconf}. Specifically, under Figure \ref{fig:DAGmediation} \textbf{(B)} and using Equation (\ref{eq:equation1w}), $ATE(Y_1) = \exp(\beta_1) \times \frac{\E_{\boldsymbol{W},\boldsymbol{{A}}}\big[\E \big\{ g_1(\boldsymbol{\tilde{A}})\mid \boldsymbol{{W}},\boldsymbol{A}, T=1\big\} \big]}{\E_{\boldsymbol{{W}},\boldsymbol{{A}}}\big[\E \big\{ g_1(\boldsymbol{\tilde{A}})\mid \boldsymbol{{W}},\boldsymbol{A}, T=0\big\} \big]}$, $NIE(Y_1, T = 1) = \frac{\E_{\boldsymbol{W},\boldsymbol{{A}}}\big[\E \big\{ g_1(\boldsymbol{\tilde{A}})\mid \boldsymbol{{W}},\boldsymbol{A}, T=1\big\} \big]}{\E_{\boldsymbol{{W}},\boldsymbol{{A}}}\big[\E \big\{ g_1(\boldsymbol{\tilde{A}})\mid \boldsymbol{{W}},\boldsymbol{A}, T=0\big\} \big]}$, and  $NDE(Y_1,$ $T = 0) = \exp(\beta_1)$.

But again, neither $\boldsymbol{{A}}$ nor $\boldsymbol{\tilde{A}}$ are measured in practice. One could regress $Y_1$ on $T$ and $\boldsymbol{W}$, but the quantity estimated in practice would differ from $ATE(Y_1)$, $NIE(Y_1,T=1)$, and $NDE(Y_1, T = 0)$. However, under Assumption A$_1$ and Assumptions A$_2$ and A$_3$ (given below), further using the estimate from regressing $Y_2$ on $T$ and $\boldsymbol{W}$, as in method Joint-Reg proposed by Etievant et al. \cite{etievantsampson2023}, allows to ``correct'' that of $Y_1$ so that one estimates $NDE(Y_1, T = 0)$. Again, a variance estimate can be obtained from the sandwich formula based on the joint estimating equation; see Supplementary Material \ref{appendix:estimationinferenceW} for details. 
\begin{eqnarray*}
\text{A}_2: && \E \big\{ g_1(\boldsymbol{\tilde{A}})\mid \boldsymbol{W} = \boldsymbol{w},T=t\big\} \text{ can be decomposed as } \exp\{f(t) + l(\boldsymbol{w})\} \text{ for } t\in \{0,1\} \\
&& \text{ and all possible values } \boldsymbol{w} \text{ of } \boldsymbol{W}.\end{eqnarray*}
\begin{eqnarray*}
\text{A}_3: && \text{ The parametric functions of  } \boldsymbol{W} \text{ in } \E(Y_{1} \mid \boldsymbol{W}, T) \text{ and } \E(Y_{2} \mid \boldsymbol{W}, T) \text{ are correctly}\\
&& \text{specified.}
\end{eqnarray*}
If $\boldsymbol{W}$ indicates strata, only Assumptions A$_1$ and A$_2$ would be needed to apply Mantel Haenszel estimation jointly to $\{T,Y_1\}$ and $\{T,Y_2\}$, as initially proposed by Etievant et al. \cite{etievantsampson2023} with their method Joint-MH, and then obtain an unbiased estimate of $NDE(Y_1, T = 0)$; see Supplementary Material \ref{appendix:estimationinferenceW}. Finally if $\boldsymbol{W}$ only has a few large strata, one could instead apply the approach in Section \ref{subsec:nontargeted} separately in each stratum of $\boldsymbol{W}$ and then combine the stratum-specific estimates. The sufficient assumption are less stringent then, as one only needs Assumptions A$_1$ to hold separately in each stratum of $\boldsymbol{W}$.

\section{Simulation}\label{sec:simul}

In this Section, a simulation is performed to verify if, under the observational setting of Figure \ref{fig:DAGmediation} \textbf{(B)}, using $Y_2$ removes both confounding bias due to $\boldsymbol{A}$ and the portion of the causal effect of $T$ on $Y_1$ that is mediated through $\boldsymbol{\tilde{A}}$. Various scenarios are considered, defined by the models described below and with parameter values as listed in the Supplementary Tables in Supplementary Material \ref{appendix:paramval}. Note, many parameters are chosen as in \cite{etievantsampson2023}. 

For each scenario, 5,000 studies with $n \in \{5,000, 10,000\}$ subjects are generated under the setting of Figure \ref{fig:DAGmediation} \textbf{(B)} with bivariate $\boldsymbol{W}$ and univariate $A$. As in \cite{etievantsampson2023}, $\boldsymbol{W} = (W_{\text{site}}, W_{\text{age}})$ represents geographic location and age, with $W_{\text{site}}$ that can take values $w_{\text{site}} \in \Omega_{W_{\text{site}}} = \{0,1,2\}$ each with probability $\frac{1}{3}$, and $ W_{\text{age}}$ that can take values $w_{\text{age}} \in \Omega_{W_{\text{age}}} = \{15,15.5,16,\dots,20.5,21\}$, with slightly different probabilities in each of the 3 strata region; see Supplementary Table \ref{webtable:Wage}. The unobserved confounder, $A$, can take values $a \in \Omega_A =\{a_{\text{low}}, a_{\text{medium}}, a_{\text{high}}\}$ with different probabilities depending on $\boldsymbol{W}$; see Supplementary Tables \ref{webtable:Ahigh} to \ref{webtable:Alow}. Then, $T$ is generated such that $\PP(T = 1 \mid \boldsymbol{W}, A) = \frac{\exp(\alpha_T + \gamma_T W_{\text{site}} + \delta_T W_{\text{age}} + \zeta_T A)}{1+\exp(\alpha_T + \gamma_T W_{\text{site}} + \delta_T W_{\text{age}} + \zeta_T A)}$; see Supplementary Table \ref{webtable:T} for parameter values. The unobser-ved mediator, $\tilde{A}$, can take values $\tilde{a} \in \Omega_{\tilde{A}} = \Omega_A$ with different probabilities depending on $W_{\text{age}}, A$ and $T$. More precisely, $\tilde{A} = A$ when $T = 0$, i.e., unvaccinated women do not engage in riskier behavior; see Supplementary Tables \ref{webtable:Atildahigh} to \ref{webtable:Atildalow} for P$(\tilde{A} \mid W_{\text{age}}, A, T= 1)$. $Y_1$ and the $Y_2^{(j)}$ are generated as Bernoulli with $\E(Y_1 \mid \boldsymbol{W}, T, \tilde{A})= \tilde{A} \times \exp\{\alpha_1 + \beta_1 T + \lambda_1 W_{\text{age}} + \sum_{w \in \Omega_{W_{\text{site}}}} \mu_w I(W_{\text{site}} = w)\}$ and $\E(Y_2^{(j)} \mid \boldsymbol{W}, T, \tilde{A})= \tilde{A} \times \exp\{\alpha_2^{(j)} + \mu_2^{(j)} W_{\text{site}} + \lambda_2^{(j)} W_{\text{age}}\}$, where $I()$ is the Indicator function, $j \in \{1, \dots, N_{NT}\}$. $\alpha_1$ and the $\alpha_2^{(j)}$ are chosen such that $\PP(Y_1 = 1) = 0.14$, $0.05$, or $0.025$, and the $\PP(Y_2^{(j)} = 1)$ are fixed to the values given in Supplementary Table \ref{webtable:Y2}; see Supplementary Table \ref{webtable:Y1} for other parameter values. Finally, $Y_2$ is computed as $ \sum_{j=1}^{N_{NT}} Y_{2}^{(j)}$. 

Methods Joint-MH and Joint-Reg that use $Y_2$ are considered, as proposed by Etievant et al. \cite{etievantsampson2023} when handling continuous $\boldsymbol{W}$ or categorical $\boldsymbol{W}$ with many strata; also see Supplementary Material \ref{appendix:estimationinferenceW}. The standard errors are estimated from the corresponding sandwich formulas. Using the true value of $\log\{NDE(Y_1, T = 0)\}$, the mean relative bias over the 5,000 studies is computed in each scenario. As a comparison, estimation with the ``naive'' methods MH and Reg that do not use $Y_2$ is also considered. Finally, the mean sandwich standard error and the sample standard deviation over the 5,000 studies are computed in each scenario and for each method. The R functions implementing the methods that had been made available by Etievant et al. \cite{etievantsampson2023} are used. For method Joint-MH, one could also use the NegativeControlOutcomeAdjustment R package available on \href{https://cran.rproject.org/web/packages/
NegativeControlOutcomeAdjustment/index.html}{CRAN}. For Reg and Joint-Reg, $W_{\text{age}}$ is assumed to affect $Y_1$ and $Y_2$ through linear quadratic functions. Note, Assumption A$_1$ holds under this simulation setting, but neither A$_2$ nor A$_3$ do.

The results for the scenarios with $n = 10, 000$ are given in Table \ref{table:conf}. Methods Joint-MH and Joint-Reg allow estimation of the log-natural direct effect of the vaccine on the targeted outcome with very small bias (columns 1 and 2), and the corresponding sandwich formulas allow appropriate variance estimation (columns 5 to 8). Joint-MH has slightly less bias than Joint-Reg, because it does not require Assumption A$_3$. Even though Assumptions A$_2$ and A$_3$ are not met, both methods work well. In comparison, methods MH and Reg yield strongly biased estimates (columns 3 and 4). Similar comments can be made for studies with $n = 5, 000$ subjects; see Supplementary Table \ref{webtable:confn5000} in Supplementary Material \ref{appendix:simuladditionalres}. The boxplots of the estimates obtained with the different methods are also displayed in Figure \ref{fig:conf}. As expected, the estimates obtained with methods Joint-MH or Joint-Reg are much closer to $\log\{NDE(Y_1, T=0)\}$ than to $\log\{ATE(Y_1)\}$. Two remarks follow. First, $\log\{NDE(Y_1, T=0)\} < \log\{ATE(Y_1)\} < 0$ because the vaccine is protective ($\beta_1 <0$, see Supplementary Table \ref{webtable:Y1}) and vaccinated women tend to engage in riskier behavior after vaccination (see Supplementary Tables \ref{webtable:Atildahigh} to \ref{webtable:Atildalow}). As a reference, relative biases with respect to $\log\{ATE(Y_1)\}$ are given in Supplementary Table \ref{webtable:biasATE} in Supplementary Material \ref{appendix:simuladditionalres}. Then, because women engaging in riskier sexual behavior are more likely to get the vaccine ($\zeta_T = 1$, see Supplementary Table \ref{webtable:T}), the confounding bias leads to an underestimated protective effect of the vaccine. Methods Joint-MH and Joint-Reg show similar performances in an additional simulation where the direction of the confounding bias is opposite; see Supplementary Material \ref{webapp:additionalsimul}.

\section{Discussion}\label{sec:discussion}

Etievant et al. \cite{etievantsampson2023} used non-targeted HPV infections as a NCO when estimating vaccine effects on targeted strains from observational data, to remove or reduce confounding bias due to unmeasured factors such as sexual behavior. But risk compensation could also occur, as certain women could engage in riskier sexual behaviors if they think they are protected after vaccination. In such a setting, infections with non-targeted HPV strains are not a valid NCO as vaccination has an effect on them through the change of sexual behavior. This work investigated whether the approach proposed by Etievant et al. \cite{etievantsampson2023} would still lead to a causally meaningful quantity in the presence of risk compensation. It was found that under similar assumptions as in \cite{etievantsampson2023}, using non-targeted HPV infections can remove both confounding bias and the portion of the causal effect that is mediated through the change of behavior (Sections \ref{subsec:nontargeted} and \ref{subsec:conf}). In other words, one assesses the direct immunological effect of the vaccine on the targeted infection.

But while this quantity has a clear causal meaning, one could argue its relevance is limited from a public health perspective as it would probably suggest higher protection from the vaccine than what women would effectively experience. Unfortunately, the total (i.e., immunological and behavioral) causal effect of the vaccine cannot be estimated because of unmeasured confounding due to sexual behavior pre-vaccination. An unblinded RCT would be a way to assess the vaccine effect on the targeted strains while accounting for the possible change of behavior. In addition, infections with non-targeted HPV strains could then be utilized to quantify risk compensation; see Supplementary Materials \ref{sec:connection} and \ref{appendix:simulunblind}. Although unblinded RCTs often come with limitations \citep{xu2024unblindchallenge}, this design could be further investigated for this purpose, especially since blinding is impossible for exposures such as circumcision and for which risk compensation is of concern \citep{agot2007circumcision, white2008circumcision}.

The current approach is based on the conditional mean models in Equations (\ref{eq:equation1}) and (\ref{eq:equation2}) (Section \ref{subsec:notation} and Supplementary Material \ref{appendix:models}). Assumption A$_1$, needed for appropriate analyses, states that $\boldsymbol{\tilde{A}}$ affects targeted and non-targeted strains proportionally (Section \ref{subsec:nontargeted} and Supplementary Material \ref{sec:connection}). This is plausible for factors related to sexual behavior, as the targeted and non-targeted strains share the same sexual transmission route. More assumptions are needed to estimate a causally meaningful quantity in the presence of a continuous observed confounder, or if it is categorical with many strata (Section \ref{subsec:conf} and Supplementary Material \ref{appendix:outcomes}). As in \cite{etievantsampson2023}, sufficient assumptions A$_2$ and A$_3$ are strong and realistically cannot be expected to hold in practice. Nevertheless, encouraging results were obtained in the simulation where assumptions other than A$_1$ were violated (Section \ref{sec:simul}). 

The method proposed by Etievant et al. \cite{etievantsampson2023} relies on several assumptions, but the rationale originated from using a causal perspective. Dema et al. \cite{dema2025} also proposed using non-targeted HPV infections to control for differences in sexual behavior between vaccinated and unvaccinated women. However, the causal interpretation of their estimate was not investigated. Future work could explore this point (see Supplementary Material \ref{appendix:rationale} for preliminary remarks).

In conclusion, vaccine efficacy estimates obtained with the methods of Etievant et al. \cite{etievantsampson2023} may be valid under other settings than the ones they initially considered. This work further emphasizes the usefulness of recording cervical infections with non-targeted strains when performing HPV genotyping in efficacy studies, to potentially improve inference on targeted strains and better understand the mechanisms of actions of the vaccine.

In this work the focus was on HPV vaccines, but there could be other applications. For example, risk compensation after Covid vaccination has attracted attention \citep{ioannidis2021covid, mccoll2024covid, smart2024covid}, and because influenza shares the same transmission route as Covid \citep{luo2023covidflu} it could potentially be used in a similar fashion as non-targeted HPV infections.

\section*{Declarations}

\subsection*{Acknowledgments}

The author is grateful to Dean Follmann at the National Institute of Allergy and Infectious Diseases, and to the Editor and anonymous Reviewers for insightful comments on the preliminary version of this article. This work utilized the computational resources of the \href{https://gricad.univ-grenoble-alpes.fr/}{GRICAD HPC platforms}. 

\subsection*{Research ethics}

Not applicable.

\subsection*{Informed consent}

Not applicable.

\subsection*{Author contributions}

Original idea, L.E. ; Conceptualization, L.E.; Original draft preparation, L.E.; Draft review and editing, L.E. The author has accepted responsibility for the entire content of this manuscript and approved its submission.

\subsection*{Use of Large Language Models, AI and Machine Learning Tools}

None declared.

\subsection*{Conflict of Interest}

The author states no conflict of interest.

\subsection*{Research funding}

None declared.

\subsection*{Data Availability}

No data pertain to this work. R code and functions used for the simulation studies are openly available on GitHub at \href{https://github.com/Etievant/RiskCompensationNonTargetedHPV}{https://github.com/Etievant/RiskCompensationNonTargetedHPV}.

\subsection*{Supplementary Material}

This article contains supplementary material.

\bibliography{biblio}

\newpage

\begin{landscape}
\begin{table}[ht]
\centering
\begin{tabular}{ccccccccccc}
\hline
\hline
\multicolumn{4}{c}{\multirow{2}{*}{Relative bias}}  & \multicolumn{2}{c}{Sample standard} & \multicolumn{2}{c}{Mean sandwich} &  \\
\multirow{2}{*}{Joint-MH} & \multirow{2}{*}{Joint-Reg} & \multirow{2}{*}{MH} & \multirow{2}{*}{Reg} & \multicolumn{2}{c}{deviation} & \multicolumn{2}{c}{standard error} & P($Y_1 = 1$) & ($a_{\text{low}}, a_{\text{medium}}, a_{\text{high}}$) & corr($Y_1,Y_2$) \\ 
 & & & & Joint-MH & Joint-Reg & Joint-MH & Joint-Reg &  \\ 
  \hline
0.024 & 0.026 & 1.230 & 1.224 & 0.060 & 0.058 & 0.060 & 0.059 & 0.14 & (0,1,2.5) & 0.250 \\ 
  0.023 & 0.030 & 0.974 & 0.973 & 0.059 & 0.058 & 0.059 & 0.058 & 0.14 & (0,1,2) & 0.224 \\ 
  0.017 & 0.024 & 0.786 & 0.788 & 0.060 & 0.058 & 0.059 & 0.058 & 0.14 & (0,0.75,1.5) & 0.226 \\ 
  0.028 & 0.034 & 1.235 & 1.233 & 0.105 & 0.103 & 0.105 & 0.103 & 0.05 & (0,1,2.5) & 0.142 \\ 
  0.025 & 0.035 & 0.975 & 0.978 & 0.105 & 0.102 & 0.104 & 0.102 & 0.05 & (0,1,2) & 0.127 \\ 
  0.015 & 0.024 & 0.784 & 0.788 & 0.105 & 0.103 & 0.104 & 0.102 & 0.05 & (0,0.75,1.5) & 0.128 \\ 
  0.029 & 0.036 & 1.235 & 1.235 & 0.151 & 0.148 & 0.150 & 0.148 & 0.025 & (0,1,2.5) & 0.099 \\ 
  0.026 & 0.037 & 0.977 & 0.980 & 0.150 & 0.146 & 0.149 & 0.146 & 0.025 & (0,1,2) & 0.089 \\ 
  0.014 & 0.024 & 0.783 & 0.788 & 0.149 & 0.147 & 0.149 & 0.147 & 0.025 & (0,0.75,1.5) & 0.090 \\ 
   \hline
   \hline
\end{tabular}
\caption{Mean relative bias, sample standard deviation, and mean sandwich standard error, for the estimates proposed by Etievant et al. \cite{etievantsampson2023} over 5,000 simulated studies under the observational setting of Figure \ref{fig:DAGmediation} \textbf{(B)} described in Section \ref{sec:simul}, with $n = $ 10,000. Relative bias is computed with respect to the true log-natural direct effect of the vaccine on the targeted outcome. Methods Joint-MH and Joint-Reg use targeted and non-targeted infections. As a comparison, the mean relative bias with methods MH and Reg that do not use non-targeted infections is also shown.}\label{table:conf}
\end{table}
\end{landscape}

\begin{figure}[ht]
\begin{minipage}[c]{0.5\linewidth}
\centering
\includegraphics[width=0.8\linewidth]{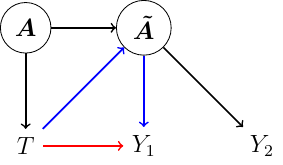}
\end{minipage}\hfill
\begin{minipage}[c]{0.5\linewidth}
\centering
\includegraphics[width=0.8\linewidth]{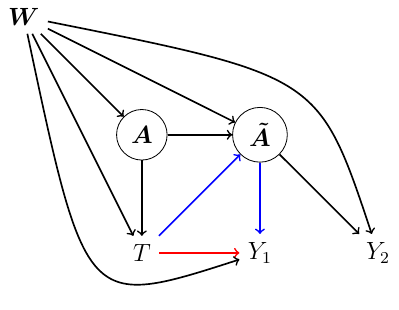}
\end{minipage}

\vspace{0.4cm}

\begin{minipage}[c]{0.5\linewidth}
\centering
\textbf{(A)}
\end{minipage}\hfill
\begin{minipage}[c]{0.5\linewidth}
\centering
\textbf{(B)}
\end{minipage}

\caption{Graph depicting the relationships among the variables in the observational setting - \textbf{(A)} Without measured confounders. - \textbf{(B)} With measured confounders. Circled variables are unmeasured. Typically, $\boldsymbol{W}$ contains geographic location and age, $\boldsymbol{A}$ is sexual behavior pre-vaccination, $\boldsymbol{\tilde A}$ is sexual behavior post-vaccination, $T$ is HPV vaccination status, $Y_{1}$ is the targeted HPV infection outcome, and $Y_{2}$ is the non-targeted infection HPV outcome.}\label{fig:DAGmediation}
\end{figure}

\newpage

\begin{figure}[ht]
    \centering
    \includegraphics[width=1\linewidth]{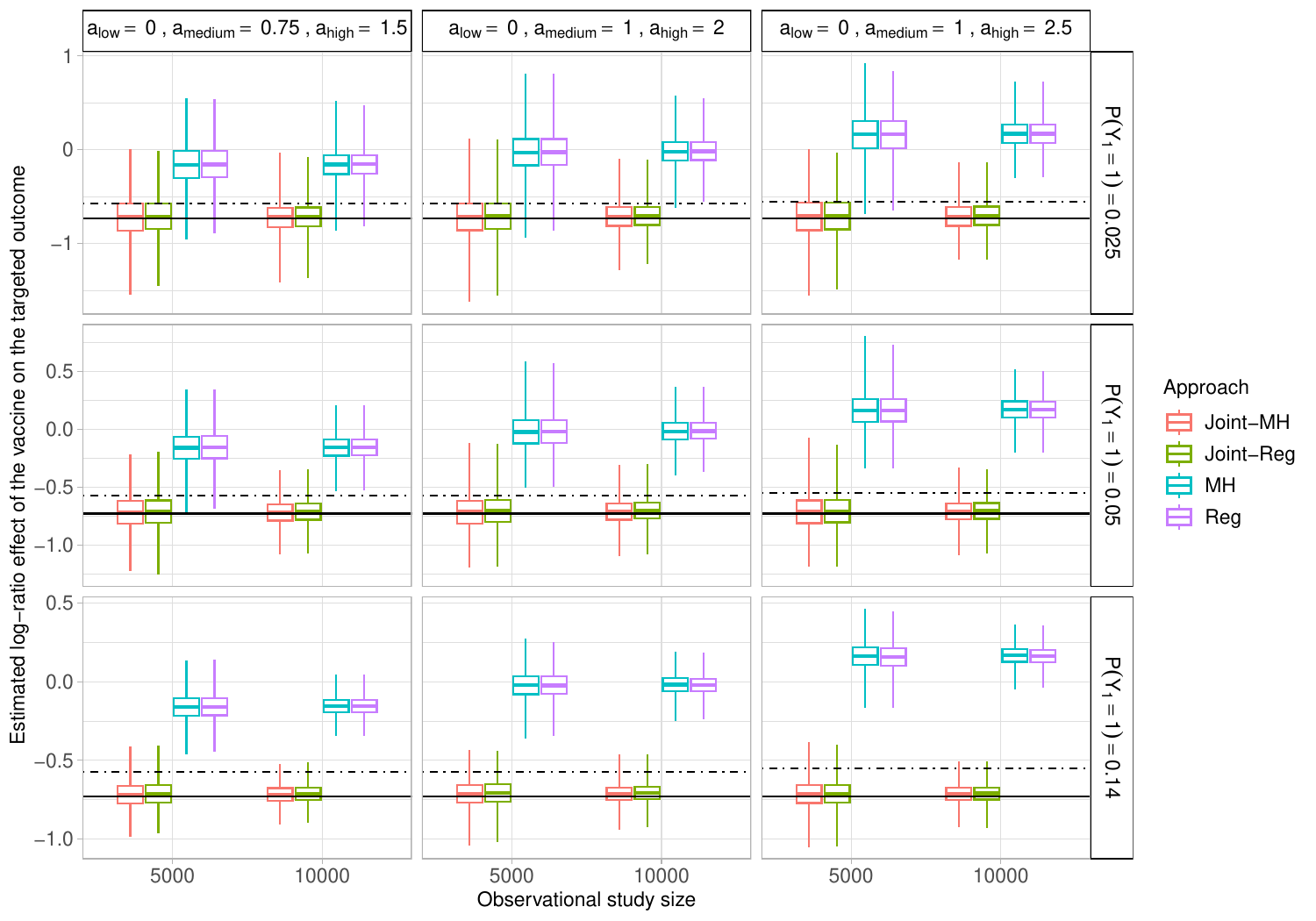}
    \caption{Boxplots of estimates with Joint-MH (red), Joint-Reg (green), MH (blue), and Reg (purple), from 5,000 simulated studies under the observational setting of Figure \ref{fig:DAGmediation} \textbf{(B)} described in Section \ref{sec:simul} and in various scenarios. The true log-natural direct effect of $T$ on $Y_1$ is indicated by the horizontal black line, while the true log-average total effect is indicated by the dash-dotted line.} \label{fig:conf}
\end{figure}

\newpage

\section*{Alternative text describing the figures}

Figure \ref{fig:DAGmediation} (A): An arrow goes from $\boldsymbol{{A}}$ to $T$, and an arrow goes from $\boldsymbol{{A}}$ to $\boldsymbol{\tilde{A}}$. A blue arrow goes from $T$ to $\boldsymbol{\tilde{A}}$, and a red arrow goes from $T$ to $Y_1$. A blue arrow goes from $\boldsymbol{\tilde{A}}$ to $Y_1$, and an arrow goes from $\boldsymbol{\tilde{A}}$ to $Y_2$. $\boldsymbol{{A}}$ and $\boldsymbol{\tilde{A}}$ are circled.\\[-0.4cm]

\noindent Figure \ref{fig:DAGmediation} (B): Five arrows go from $\boldsymbol{{W}}$ to $\boldsymbol{{A}}$, $T$, $\boldsymbol{\tilde{A}}$, $Y_1$, and $Y_2$. An arrow goes from $\boldsymbol{{A}}$ to $T$, and an arrow goes from $\boldsymbol{{A}}$ to $\boldsymbol{\tilde{A}}$. A blue arrow goes from $T$ to $\boldsymbol{\tilde{A}}$, and a red arrow goes from $T$ to $Y_1$. A blue arrow goes from $\boldsymbol{\tilde{A}}$ to $Y_1$, and an arrow goes from $\boldsymbol{\tilde{A}}$ to $Y_2$. $\boldsymbol{{A}}$ and $\boldsymbol{\tilde{A}}$ are circled.\\[-0.4cm]

\noindent Figure \ref{fig:conf}: The figure contains nine panels, arranged in three columns and three rows. There is one column for the scenarios with $a_{\text{low}} = 0, a_{\text{medium}} = 0.75, a_{\text{high}} = 1.5$, one for the scenarios with $a_{\text{low}} = 0, a_{\text{medium}} = 1, a_{\text{high}} = 2$, and one for the scenarios with $a_{\text{low}} = 0, a_{\text{medium}} = 1, a_{\text{high}} = 2.5$, and there are three rows for the scenarios with $\PP(Y_1 = 1) = 0.14$, $\PP(Y_1 = 1) = 0.05$, and $\PP(Y_1 = 1) = 0.025$, respectively. The observational study size $n$ is on the $x$-axis, and the estimated log-ratio effect of the vaccine on the targeted outcome is on the $y$-axis. Each panel contains the boxplots of the 5,000 estimates with method Joint-MH in red, one for $n= 5,000$ and one for $n=10,000$, and similarly the two boxplots with method Joint-Reg in green, with method MH in cyan, and with method Reg in purple. The black horizontal solid line at $y=-0.73$ shows the true log-natural direct effect of $T$ on $Y_1$, and the black horizontal dash-dotted line at $y \approx -0.553$ or $y\approx -0.572$ shows the true log-average total effect of $T$ on $Y_1$. The log-natural direct effect of $T$ on $Y_1$ is slightly underestimated with methods Joint-MH and Joint-Reg. It is heavily underestimated with methods MH and Reg, that even suggest a protective effect of the vaccine in the scenarios in the third column.

\newpage

\thispagestyle{empty}

\LARGE 
\begin{center}
Supplementary Material for Using NonTargeted HPV Infections in Studies with Risk Compensation
\end{center}

\large
\begin{center}
    Lola Etiévant$^{1}$
\end{center}

\normalsize

\noindent $^{1}$ Univ. Grenoble Alpes, CNRS, LJK, 38000 Grenoble, France.\\

\noindent SVH Team, DATA Department, Jean Kuntzmann Laboratory, Bâtiment IMAG, 150 place du Torrent, 38401 Campus Universitaire de Saint-Martin-d'Hères, FRANCE\\

\noindent \href{mailto:lola.etievant@univ-grenoble-alpes.fr}{lola.etievant@univ-grenoble-alpes.fr} \quad \quad \href{https://orcid.org/0000-0001-7562-3550}{ORCID 0000-0001-7562-3550}

\thispagestyle{empty}

\newpage

\appendix

\addcontentsline{toc}{section}{Supplementary Material}

\section{Identifiability of the causal effects under the observa-tional setting without observed confounders}\label{appendix:identifiability}

Under the observational setting of Figure \ref{fig:DAGmediation} {(A)} in the Main Document, the following identifiability conditions hold, for any values $t$, $t^*$, and $\boldsymbol{\tilde{a}}$:
\begin{eqnarray*}
   & \text{\tiny{(Consistency assumption)}} \quad & \text{If } T=t \text{, then }Y_1^{t} = Y_1 \text{ and } \tilde{A}^{t} = \tilde{A},\\
   & \text{\tiny{(Ignorability assumption)}} \quad & Y_1^{t} \indep T \mid \boldsymbol{{A}}, \\
   & \text{\tiny{(Assumption of no confounder for the $\boldsymbol{\tilde{A}}-Y_1$ relationship that are caused by $T$)}} \quad  & Y_{1}^{t,\boldsymbol{\tilde{a}}} \indep \boldsymbol{\tilde{A}}^{t^*},\\
   & \text{\tiny{(Assumption of no unmeasured confounder for the $T-Y1$ relationship)}}  \quad &Y_{1}^{t,\boldsymbol{\tilde{a}}} \indep T,\\
   & \text{\tiny{(Assumption of no unmeasured confounder for the $\boldsymbol{\tilde{A}}-Y_1$ relationship)}}  \quad &Y_{1}^{t,\boldsymbol{\tilde{a}}} \indep  \boldsymbol{\tilde{A}} \mid T,\\
   & \text{\tiny{(Assumption of no unmeasured confounder for the $T-\boldsymbol{\tilde{A}}$ relationship}}  \quad &\boldsymbol{\tilde{A}}^{t^*} \indep T \mid \boldsymbol{{A}}.
\end{eqnarray*}
In this way, 
\begin{eqnarray*}
    \E \left( Y_1^{t} \right) &=& \E_{\boldsymbol{A}} \big\{ \E\left( Y_1 \mid \boldsymbol{{A}}, T=t \right)\big\}, \\
    &=& \E_{\boldsymbol{{A}}}\big[\E_{\boldsymbol{\tilde{A}}} \big\{ \E\big( Y_1 \mid T=t, \boldsymbol{{\tilde{A}}} \big)\mid \boldsymbol{A}, T=t\big\} \big],
\end{eqnarray*}
where the second equality comes from the fact that $ Y_1 \indep \boldsymbol{{A}} \mid \{T, \boldsymbol{\tilde{A}} \}$, and $$\E \left(Y_{1}^{t,\boldsymbol{\tilde{A}}^{t^*}} \right) = \E_{\boldsymbol{{A}}}\big[\E_{\boldsymbol{\tilde{A}}} \big\{ \E\big( Y_1 \mid T=t, \boldsymbol{{\tilde{A}}} \big)\mid \boldsymbol{A}, T=t^*\big\} \big].$$

\section{Estimation and inference with data on $\{T,Y_1, Y_2 \}$ under the observational setting without observed confounders}\label{appendix:estimationinference}

The setting in Figure \ref{fig:DAGmediation} {(A)} in the Main Document is considered, with the estimation technique proposed by Etievant et al. \cite{etievantsampson2023} when $\boldsymbol{W}=\emptyset$. First note that using Equations (\ref{eq:equation1}) and (\ref{eq:equation2}) in Section \ref{subsec:notation} in the Main Document, one can write
\begin{eqnarray}
\E(Y_{1} \mid T) &=& \exp(\alpha_1^*+ \beta_1^* T), \label{eq:equation1star}    \\
 \text{ and } \hspace{4.5cm} \E(Y_{2} \mid T) &=& \exp(\alpha_2^*+ \beta_2^* T),  \hspace{5.1cm}\label{eq:equation2star}  
\end{eqnarray}
with $\alpha_1^* = \alpha_1 + \log\big[ \E \big\{ g_1(\boldsymbol{\tilde{A}})\mid T=0\big\}\big]$ and $\beta_1^* = \beta_1 + \log\left[\frac{\E \big\{ g_1(\boldsymbol{\tilde{A}})\mid T=1\big\}}{\E \big\{ g_1(\boldsymbol{\tilde{A}})\mid T=0\big\}}\right]$, and with $\alpha_2^* = \alpha_2 + \log\big[ \E \big\{ g_2(\boldsymbol{\tilde{A}})\mid T=0\big\}\big]$ and $\beta_2^* = \log\left[\frac{\E \big\{ g_2(\boldsymbol{\tilde{A}})\mid T=1\big\}}{\E \big\{ g_2(\boldsymbol{\tilde{A}})\mid T=0\big\}}\right]$. 
Then observe that under Assumption A$_1$ given in Section \ref{subsec:nontargeted} in the Main Document and recalled below, $\beta_1^* - \beta_2^* = \beta_1$. 
$$\text{A}_1: \text{ There exists }k > 0 \text{ such that } g_2(\boldsymbol{\tilde{a}}) = k \times g_1(\boldsymbol{\tilde{a}}) \text{ for all possible values } \boldsymbol{\tilde{a}} \text{ of } \boldsymbol{\tilde{A}}.$$
The standard approach using data on $\{T,Y_1\}$ would simply estimate parameter $\beta_1^*$. On the other hand, the approach proposed by Etievant et al. \cite{etievantsampson2023} (the joint approach with no covariates, Joint-NC) uses data on $\{T,Y_1,Y_2\}$ and consists in jointly estimating $\frac{\E(Y_{1} \mid T = 1)}{\E(Y_{1} \mid T = 0)}$ and $\frac{\E(Y_{2} \mid T = 1)}{\E(Y_{2} \mid T = 0)}$. As a reminder, this can be done by solving for $\boldsymbol{\theta} = (\alpha_1^*, \beta_1^*, \alpha_2^*,$ $\beta_2^*)$ the joint estimating equation $\sum_{i=1}^n \boldsymbol{U}(Y_{1,i}, Y_{2,i}, T_i; \boldsymbol{\theta}) = 0$, with
$$
    \boldsymbol{U}(Y_{1,i}, Y_{2,i}, T_i; \boldsymbol{\theta}) = \begin{pmatrix}
        \frac{Y_{1,i} - p_{1,i}^*}{1 - p_{1,i}^*} \\
        \frac{T_i(Y_{1,i} - p_{1,i}^*)}{1 - p_{1,i}^*} \\
        Y_{2,i} - p_{2,i}^* \\
        T_i(Y_{2,i} - p_{2,i}^*)
    \end{pmatrix},
$$
and where $p_{1,i}^* = \exp(\alpha_1^* + \beta_1^* T_i)$, and $p_{2,i}^* = \exp(\alpha_2^* + \beta_2^* T_i)$. This estimating equation is obtained from the log-likelihoods of the models in Supplementary Equations (\ref{eq:equation1star}) and (\ref{eq:equation2star}). It leads to $\hat \alpha_1^* = \log\left\{\frac{\sum_{i=1}^{n} (1 - T_i) Y_{1,i}}{\sum_{i=1}^{n}(1 - T_i)} \right\}$, $\hat \beta_1^* = \log \left\{\frac{\sum_{i=1}^{n} T_i Y_{1,i}}{\sum_{i=1}^{n} T_i} \times \frac{\sum_{i=1}^{n}(1 - T_i)}{\sum_{i=1}^{n} (1 - T_i) Y_{1,i}} \right\}$, $\hat \alpha_2^* = \log\left\{\frac{\sum_{i=1}^{n} (1 - T_i) Y_{2,i}}{\sum_{i=1}^{n}(1 - T_i)} \right\}$ and $\hat \beta_2^* = \log \left\{\frac{\sum_{i=1}^{n} T_i Y_{2,i}}{\sum_{i=1}^{n} T_i} \times \right.$ $\left.\frac{\sum_{i=1}^{n}(1 - T_i)}{\sum_{i=1}^{n} (1 - T_i) Y_{2,i}} \right\}$. The four variances and six covariances can then be estimated from the sandwich formula
\begin{eqnarray*}
\frac{1}{n} \left\{ \frac{1}{n} \sum_{i=1}^{n} \frac{\partial \boldsymbol{U}(Y_{1,i}, Y_{2,i}, T_{i}; {\boldsymbol{\theta}})}{\partial \boldsymbol{\theta}}_{|\boldsymbol{\theta} = \widehat{\boldsymbol{\theta}}} \right\}^{-1} \hspace{-0.7cm}
&& \left[ \frac{1}{n} \sum_{i=1}^{n} \left\{ \boldsymbol{U}(Y_{1,i}, Y_{2,i}, T_{i}; \widehat{\boldsymbol{\theta}}) \right\} \left\{ \boldsymbol{U}(Y_{1,i}, Y_{2,i}, T_{i}; \widehat{\boldsymbol{\theta}}) \right\}^\top \right] \nonumber\\
&& \left[ \left\{ \frac{1}{n} \sum_{i=1}^{n} \frac{\partial \boldsymbol{U}(Y_{1,i}, Y_{2,i}, T_{i}; {\boldsymbol{\theta}})}{\partial \boldsymbol{\theta}}_{|\boldsymbol{\theta} = \widehat{\boldsymbol{\theta}}} \right\}^{-1} \right]^\top \hspace{-0.1cm},\label{eq:sandwich}
\end{eqnarray*}
with
\begin{equation*}
\frac{\partial \boldsymbol{U}(Y_{1,i}, Y_{2,i}, T_i; \boldsymbol{\theta})}{\partial \boldsymbol{\theta}} = 
\begin{pmatrix}
\frac{p_{1,i}^* (Y_{1,i} - 1)}{(1 - p_{1,i}^*)^2} & \frac{T_i  p_1^* (Y_{1,i} - 1)}{(1 - p_{1,i}^*)^2} & 0 & 0 \\
\frac{T_i p_{1,i}^* (Y_{1,i} - 1)}{(1 - p_{1,i}^*)^2} & \frac{T_i p_1^* (Y_{1,i} - 1)}{(1 - p_{1,i}^*)^2} & 0 & 0 \\
0 & 0 & -p_{2,i}^* & -T_i  p_{2,i}^* \\
0 & 0 & -T_i  p_{2,i}^* & -T_i  p_{2,i}^*
\end{pmatrix}.
\end{equation*}
Confidence intervals can be computed assuming asymptotic normality \citep{ziegler2011book}. Etievant et al. \cite{etievantsampson2023} finally consider $\hat \beta_1^* - \hat \beta_2^*$, whose variance estimate, denoted $\hat v_1$, is obtained from adding the variances of $\hat \beta_1^*$ and $\hat \beta_2^*$ and subtracting two times their covariance. 

As a result, under the setting of Figure \ref{fig:DAGmediation} {(A)} in the Main Document and if Assumption A$_1$ holds, one would obtain an estimate of the log-natural direct effect, denoted $\hat \beta_1$. As a remark, if one is not interested in the log-ratio effect, it is possible to simply exponentiate to obtain $\exp\big(\hat \beta_1\big) =\frac{\sum_{i=1}^{n} T_i Y_{1,i}}{\sum_{i=1}^{n} (1 - T_i) Y_{1,i}} \times \frac{\sum_{i=1}^{n} (1 - T_i) Y_{2,i}}{\sum_{i=1}^{n} T_i Y_{2,i}}$ and the variance estimate can be obtained from the Delta Method, with $\hat v_1 \left\{\exp\big(\hat \beta_1\big)\right\}^2$ \citep{ziegler2011book}.

\newpage

\section{Identifiability of the causal effects under the observa-tional setting with observed confounders}\label{appendix:identifiabilityconf}

Under the observational setting of Figure \ref{fig:DAGmediation} {(B)} in the Main Document, the following identifiability conditions hold, for any values $t$, $t^*$, and $\boldsymbol{\tilde{a}}$:
\begin{eqnarray*}
   & \text{\tiny{(Consistency assumption)}} \quad & \text{If } T=t \text{, then }Y_1^{t} = Y_1 \text{ and } \tilde{A}^{t} = \tilde{A},\\
   & \text{\tiny{(Ignorability assumption)}} \quad & Y_1^{t} \indep T \mid \{\boldsymbol{{W}}, \boldsymbol{{A}}\}, \\
   & \text{\tiny{(Assumption of no confounder for the $\boldsymbol{\tilde{A}}-Y_1$ relationship that are caused by $T$)}} \quad  & Y_{1}^{t,\boldsymbol{\tilde{a}}} \indep \boldsymbol{\tilde{A}}^{t^*} \mid \boldsymbol{{W}},\\
   & \text{\tiny{(Assumption of no unmeasured confounder for the $T-Y1$ relationship)}}  \quad &Y_{1}^{t,\boldsymbol{\tilde{a}}} \indep T \mid \boldsymbol{{W}},\\
   & \text{\tiny{(Assumption of no unmeasured confounder for the $\boldsymbol{\tilde{A}}-Y_1$ relationship)}}  \quad &Y_{1}^{t,\boldsymbol{\tilde{a}}} \indep  \boldsymbol{\tilde{A}} \mid \{\boldsymbol{{W}}, T\} ,\\
   & \text{\tiny{(Assumption of no unmeasured confounder for the $T-\boldsymbol{\tilde{A}}$ relationship}}  \quad &\boldsymbol{\tilde{A}}^{t^*} \indep T \mid \{\boldsymbol{{W}}, \boldsymbol{{A}}\}.
\end{eqnarray*}
\normalsize In this way, 
\begin{eqnarray*}
    \E \left( Y_1^{t} \right) &=& \E_{\boldsymbol{{W}},\boldsymbol{A}} \big\{ \E\left( Y_1 \mid \boldsymbol{{W}},\boldsymbol{{A}}, T=t \right)\big\}, \\
    &=& \E_{\boldsymbol{{W}},\boldsymbol{{A}}}\big[\E_{\boldsymbol{\tilde{A}}} \big\{ \E\big( Y_1 \mid \boldsymbol{{W}}, T=t,\boldsymbol{{\tilde{A}}} \big)\mid \boldsymbol{{W}},\boldsymbol{A}, T=t\big\} \big],
\end{eqnarray*}
where the second equality comes from the fact that $ Y_1 \indep \boldsymbol{{A}} \mid \{\boldsymbol{{W}},T, \boldsymbol{\tilde{A}} \}$, and
\begin{eqnarray*}
  \E \left(Y_{1}^{t,\boldsymbol{\tilde{A}}^{t^*}} \right) &=& \E_{\boldsymbol{{W}},\boldsymbol{{A}}}\big[\E_{\boldsymbol{\tilde{A}}} \big\{ \E\big( Y_1 \mid \boldsymbol{{W}},\boldsymbol{{\tilde{A}}}, T=t \big)\mid \boldsymbol{{W}},\boldsymbol{A}, T=t^*\big\} \big].
\end{eqnarray*}

\newpage

\section{Estimation and inference with data on $\{\boldsymbol{W}, T, Y_1, Y_2 \}$ under the observational setting with observed confoun-ders} \label{appendix:estimationinferenceW}

The setting of Figure \ref{fig:DAGmediation} {(B)} in the Main Document is considered, with the estimation techniques proposed by Etievant et al. \cite{etievantsampson2023} when $\boldsymbol{W}\neq\emptyset$. First, using Equation (\ref{eq:equation1w}) in Section \ref{subsec:conf} in the Main Document and as $\E(Y_{1} \mid \boldsymbol{W}, T ) = \E_{\boldsymbol{\tilde{A}}} \big\{ \E\big( Y_1 \mid \boldsymbol{W}, T, \boldsymbol{{\tilde{A}}} \big)\mid \boldsymbol{W},T\big\}$, it follows that
\begin{eqnarray*}
\E(Y_{1} \mid \boldsymbol{W}, T ) 
&=& \exp\Bigg(\alpha_1 + \beta_1 T + s_1(\boldsymbol{W}) + \log\left[\frac{\E \big\{ g_1(\boldsymbol{\tilde{A}})\mid \boldsymbol{W},T=1\big\}}{\E \big\{ g_1(\boldsymbol{\tilde{A}})\mid \boldsymbol{W},T=0\big\}}\right]T \\
&& + \log\Big[\E \big\{ g_1(\boldsymbol{\tilde{A}})\mid \boldsymbol{W},T=0\big\}\Big] \Bigg).
\end{eqnarray*}
Similarly, using Equation (\ref{eq:equation2w}) in Section \ref{subsec:conf} in the Main Document, 
\begin{eqnarray*}
\E(Y_{2} \mid \boldsymbol{W}, T) 
&=& \exp\Bigg(\alpha_2 + s_2(\boldsymbol{W}) + \log\left[\frac{\E \big\{ g_2(\boldsymbol{\tilde{A}})\mid \boldsymbol{W},T=1\big\}}{\E \big\{ g_2(\boldsymbol{\tilde{A}})\mid \boldsymbol{W},T=0\big\}}\right]T \\
&& + \log\Big[\E \big\{ g_2(\boldsymbol{\tilde{A}})\mid \boldsymbol{W},T=0\big\}\Big] \Bigg).
\end{eqnarray*}
If Assumption A$_1$ given in Section \ref{subsec:nontargeted} in the Main Document and recalled in Supplementary Material \ref{appendix:identifiabilityconf} holds, and if $\log\left[\frac{\E \big\{ g_1(\boldsymbol{\tilde{A}})\mid \boldsymbol{W},T=1\big\}}{\E \big\{ g_1(\boldsymbol{\tilde{A}})\mid \boldsymbol{W},T=0\big\}}\right]$ is independent of $\boldsymbol{W}$, one can rewrite
\begin{eqnarray}
\E(Y_{1} \mid \boldsymbol{W}, T)=\exp\{\alpha_1^* + \beta_1^* T + s_1^*(\boldsymbol{W})\}, \label{eq:mod1wbias}
\end{eqnarray}
and 
\begin{eqnarray}
\E(Y_{2} \mid \boldsymbol{W}, T )=\exp\{\alpha_2^* + \beta_2^* T + s_2^*(\boldsymbol{W})\}, \label{eq:mod2wbias} 
\end{eqnarray}
with $\beta_1^* - \beta_2^* = \beta_1$. Following Etievant et al. \cite{etievantsampson2023}, if Assumption A$_2$ given in Section \ref{subsec:conf} in the Main Document and recalled below holds, then $\log\left[\frac{\E \big\{ g_1(\boldsymbol{\tilde{A}})\mid \boldsymbol{W},T=1\big\}}{\E \big\{ g_1(\boldsymbol{\tilde{A}})\mid \boldsymbol{W},T=0\big\}}\right]$ does not depend on $\boldsymbol{W}$. More specifically, and if Assumption A$_1$ also holds, $\alpha_1^* = \alpha_1 + f(0)$, $\alpha_2^* = \alpha_2 + f(0)$, $\beta_1^* = \beta_1 + f(1) - f(0)$, $\beta_2^* = f(1) - f(0)$, $s_1^*(\boldsymbol{w}) = s_1(\boldsymbol{w}) + l(\boldsymbol{w})$ and $s_2^*(\boldsymbol{w}) = s_2(\boldsymbol{w}) + l(\boldsymbol{w})$ for all possible value $\boldsymbol{w}$ of $\boldsymbol{W}$.
\begin{eqnarray*}
\text{A}_2: && \E \big\{ g_1(\boldsymbol{\tilde{A}})\mid \boldsymbol{W} = \boldsymbol{w},T=t\big\} \text{ can be decomposed as } \exp\{f(t) + l(\boldsymbol{w})\} \text{ for } t\in \{0,1\} \\ 
&& \text{ and all possible values } \boldsymbol{w} \text{ of } \boldsymbol{W}.
\end{eqnarray*}

The standard approach when handling a continuous $\boldsymbol{W}$ would probably be regressing $Y_1$ on $\boldsymbol{W}$ and $T$, and again this can be done by solving the estimating equation derived from the log-likelihood of the model in Supplementary Equation (\ref{eq:mod1wbias}) \citep{etievantsampson2023}. However, one would simply estimate parameter $\beta_1^*$ (assuming the parametric function $s_1^*$ had correctly been specified in the regression). On the other hand, the approach proposed by Etievant et al. \cite{etievantsampson2023} (the joint regression approach, Joint-Reg) consists in solving the joint estimating equation derived from the log-likelihoods of the models in Supplementary Equations (\ref{eq:mod1wbias}) and (\ref{eq:mod2wbias}) and then considering estimate $\hat \beta_1^* - \hat \beta_2^*$. For example, assuming that the observed confounder is univariate and that $s_1^* :w \mapsto \mu_1^* w + \nu_1^* w^2$ and $s_2^* :w \mapsto \mu_2^* w + \nu_2^* w^2$, one would solve for $\boldsymbol{\theta} = (\alpha_1^*, \beta_1^*, \mu_1^*, \nu_1^*, \alpha_2^*,$ $\beta_2^*, \mu_2^*, \nu_2^*)$ the joint estimating equation $\sum_{i=1}^n \boldsymbol{U}(Y_{1,i}, Y_{2,i}, W_i,T_i; \boldsymbol{\theta}) = 0$, with 
\begin{equation*}
    \boldsymbol{U}(Y_{1,i}, Y_{2,i}, W_i, T_i; \boldsymbol{\theta}) = 
    \begin{pmatrix}
        \frac{Y_{1,i} - p^*_{1,i}}{1 - p^*_{1,i}} \\
        \frac{T_i(Y_{1,i} - p^*_{1,i})}{1 - p^*_{1,i}} \\
        \frac{W_i(Y_{1,i} - p^*_{1,i})}{1 - p^*_{1,i}} \\
        \frac{W_i^2(Y_{1,i} - p^*_{1,i})}{1 - p^*_{1,i}} \\
        Y_{2,i} - p^*_{2,i} \\
        T_i(Y_{2,i} - p^*_{2,i}) \\
        W_i(Y_{2,i} - p^*_{2,i})\\
        W_i^2(Y_{2,i} - p^*_{2,i})
    \end{pmatrix},
\end{equation*}
and where $p_{1,i}^* = \exp(\alpha_1^* + \beta_1^* T_i + \mu_1^* W_i + \nu_1^* W_i^2)$, and $p_{2,i}^* = \exp(\alpha_2^* + \beta_2^* T_i + \mu_2^* W_i + \nu_2^* W_i^2)$. An iterative algorithm would be used to obtain $\widehat{\boldsymbol{\theta}}$, and the variances and covariances would be obtained from the sandwich formula as in Supplementary Material \ref{appendix:estimationinference}, but with 
\begin{equation*}
    \frac{\partial \boldsymbol{U}(Y_{1,i},Y_{2,i},W_i,T_i;\boldsymbol{\theta})}{\partial \boldsymbol{\theta}} = 
    \begin{pmatrix} 
    \boldsymbol{M}_1 & \boldsymbol{0}_{4\times4}\\
     \boldsymbol{0}_{4\times4} & \boldsymbol{M}_2
    \end{pmatrix},
\end{equation*}
where $\boldsymbol{0}_{4\times4}$ the null square matrix of order $4$,  
\begin{equation*}
 \boldsymbol{M}_1 = 
  \begin{pmatrix}
    \frac{p_{1,i}^* (Y_{1,i} - 1)}{(1 - p_{1,i}^*)^2} & \frac{T_i  p_1^* (Y_{1,i} - 1)}{(1 - p_{1,i}^*)^2} & \frac{W_i  p_1^* (Y_{1,i} - 1)}{(1 - p_{1,i}^*)^2} & \frac{W_i^2  p_1^* (Y_{1,i} - 1)}{(1 - p_{1,i}^*)^2} \\
    \frac{T_i  p_1^* (Y_{1,i} - 1)}{(1 - p_{1,i}^*)^2} & \frac{T_i  p_1^* (Y_{1,i} - 1)}{(1 - p_{1,i}^*)^2} & \frac{W_iT_i  p_1^* (Y_{1,i} - 1)}{(1 - p_{1,i}^*)^2} & \frac{W_i^2T_i  p_1^* (Y_{1,i} - 1)}{(1 - p_{1,i}^*)^2}\\
    \frac{W_i  p_1^* (Y_{1,i} - 1)}{(1 - p_{1,i}^*)^2} & \frac{W_i T_i  p_1^* (Y_{1,i} - 1)}{(1 - p_{1,i}^*)^2} & \frac{W_i^2  p_1^* (Y_{1,i} - 1)}{(1 - p_{1,i}^*)^2} & \frac{W_i^3  p_1^* (Y_{1,i} - 1)}{(1 - p_{1,i}^*)^2} \\
    \frac{W_i^2  p_1^* (Y_{1,i} - 1)}{(1 - p_{1,i}^*)^2} & \frac{W_i^2T_i  p_1^* (Y_{1,i} - 1)}{(1 - p_{1,i}^*)^2} & \frac{W_i^3  p_1^* (Y_{1,i} - 1)}{(1 - p_{1,i}^*)^2} & \frac{W_i^4  p_1^* (Y_{1,i} - 1)}{(1 - p_{1,i}^*)^2}
  \end{pmatrix},
\end{equation*}
and 
\begin{equation*}
 \boldsymbol{M}_2 = 
  \begin{pmatrix}
   -p_{2,i}^* & -T_i  p_{2,i}^* & -W_i  p_{2,i}^* & -W_i^2  p_{2,i}^*\\
    -T_i  p_{2,i}^* & -T_i  p_{2,i}^* & -W_i T_i  p_{2,i}^* & -W_i^2 T_i  p_{2,i}^*\\
    -W_i  p_{2,i}^* & -W_i T_i  p_{2,i}^* & -W_i^2  p_{2,i}^* & -W_i^3  p_{2,i}^*\\
   -W_i^2 p_{2,i}^* & -W_i^2 T_i  p_{2,i}^* & -W_i^3  p_{2,i}^* & -W_i^4  p_{2,i}^*
  \end{pmatrix}.
\end{equation*}

Then, if Assumption A$_1$ and Assumption A$_2$ hold under the setting of Figure \ref{fig:DAGmediation} {(B)} in the Main Document, and if $s_1^*$ and $s_2^*$ in Equations (\ref{eq:mod1wbias}) and (\ref{eq:mod2wbias}) are correctly specified (Assumption A$_3$ given in Section \ref{subsec:conf} in the Main Document), $\hat \beta_1^* - \hat \beta_2^*$ is an unbiased estimate of the log-natural direct effect.\\

Now, if $\boldsymbol{W}$ indicates strata, one can rewrite the generalized linear models in Equations (\ref{eq:equation1w}) and (\ref{eq:equation2w}) in Section \ref{subsec:conf} in the Main Document as
\begin{eqnarray}
    \E(Y_{1} \mid \boldsymbol{W},  T, \boldsymbol{\tilde{A}}) &=& g_{1}(\boldsymbol{\tilde{A}}) \exp(\alpha_{1,\boldsymbol{W}} + \beta_1 T),
    \label{eq:equation1wcat}\\
    \E(Y_{2} \mid \boldsymbol{W}, T, \boldsymbol{\tilde{A}}) &=& g_{2}(\boldsymbol{\tilde{A}}) \exp(\alpha_{2,\boldsymbol{W}}).
    \label{eq:equation2wcat}
\end{eqnarray}
Assume there are $K$ strata and $\boldsymbol{W}$ can take one of the $K$ values $\boldsymbol{w}_1, \dots, \boldsymbol{w}_k$, as in Etievant et al. \cite{etievantsampson2023}. Using similar arguments as above, if Assumption A$_{1}$ holds, and if $\log\left[\frac{\E \big\{ g_{1}(\boldsymbol{\tilde{A}})\mid \boldsymbol{W},T=1\big\}}{\E \big\{ g_{1}(\boldsymbol{\tilde{A}})\mid \boldsymbol{W},T=0\big\}}\right]$ is independent of $\boldsymbol{W}$, it follows that
\begin{eqnarray*}
\E(Y_{1} \mid \boldsymbol{W}, T)=\exp(\alpha_{1,\boldsymbol{W}}^* + \beta_1^* T), \label{eq:mod1wcatbias}\\
\E(Y_{2} \mid \boldsymbol{W}, T )=\exp(\alpha_{2,\boldsymbol{W}}^* + \beta_2^* T), \label{eq:mod2catwbias} 
\end{eqnarray*}
with $\beta_1^* - \beta_2^* = \beta_1$.
Following Etievant et al. \cite{etievantsampson2023}, if Assumption A$_{2,\boldsymbol{W}}$ given below holds, then $\log\left[\frac{\E \big\{ g_{1}(\boldsymbol{\tilde{A}})\mid \boldsymbol{W},T=1\big\}}{\E \big\{ g_{1}(\boldsymbol{\tilde{A}})\mid \boldsymbol{W},T=0\big\}}\right]$ does not depend on $\boldsymbol{W}$, and one would more specifically have $\alpha_{1,\boldsymbol{W}}^* = \alpha_{1,\boldsymbol{W}} + f(0) + l_{\boldsymbol{W}}$, $\alpha_{2,\boldsymbol{W}}^* = \alpha_{2,\boldsymbol{W}} + f(0) + l_{\boldsymbol{W}}$, $\beta_1^* = \beta_1 + f(1) - f(0)$, $\beta_2^* = f(1) - f(0)$.
\begin{eqnarray*}
\text{A}_{2,\boldsymbol{W}}: && \E \big\{ g_{1}(\boldsymbol{\tilde{A}})\mid \boldsymbol{W} = \boldsymbol{w},T=t\big\} \text{ can be decomposed as } \exp\{f(t) + l_{\boldsymbol{w}}\} \text{ with stratum-} \\ 
&& \text{specific constant } l_{\boldsymbol{w}}, \text{ for } t\in \{0,1\} \text{ and } \boldsymbol{w} \in \{\boldsymbol{w}_1, \dots, \boldsymbol{w}_k\}.
\end{eqnarray*}
Note, Assumption A$_{2,\boldsymbol{W}}$ is simply a rewriting of Assumption A$_2$ for a categorical $\boldsymbol{W}$.

\normalsize

The standard approach would probably be using Mantel-Haenszel estimation on $\{\boldsymbol{W},$ $T, Y_1\}$ \citep{tarone1981MH}, but again one would then simply estimate $\beta_1^*$. On the other hand, the approach proposed by Etievant et al. \cite{etievantsampson2023} consists in jointly estimating $\beta_1^*$ and $\beta_2^*$ with the Mantel-Haenszel (MH) method, and then considering estimate $\beta_1^*-\beta_2^*$ (the joint Mantel-Haenszel approach, Joint-MH). Again, following Etievant et al. \cite{etievantsampson2023}, this can be done by solving for $\boldsymbol{\theta} = (\beta_1^*,\beta_2^*)$ the joint estimating equation $\sum_{k=1}^{K} \omega_k \, \boldsymbol{U}\big(X_{1,k}, Z_{1,k}, X_{2,k},$ $ Z_{2,k}; \boldsymbol{\theta} \big) = 0$ with 
\begin{equation*}
    \boldsymbol{U}(X_{1,k}, Z_{1,k}, X_{2,k}, Z_{2,k};\boldsymbol{\theta}) = 
    \begin{pmatrix} 
        \frac{X_{1,k}}{n_{1,k}} - \exp(\beta_1^*) \frac{Z_{1,k}}{n_{0,k}} \\
        \frac{X_{2,k}}{n_{1,k}} - \exp(\beta_2^*) \frac{Z_{2,k}}{n_{0,k}} 
    \end{pmatrix},
\end{equation*}
and $\omega_k =\frac{n_{1,k} n_{0,k}}{n_k}$, and where $X_{1,k} = \sum_{i=1}^{n} T_i Y_{1,i} I(\boldsymbol{W}_i = \boldsymbol{w}_k)$, $Z_{1,k} = \sum_{i=1}^{n} (1 - T_i) Y_{1,i} I(\boldsymbol{W}_i = \boldsymbol{w}_k)$, $X_{2,k} = \sum_{i=1}^{n} T_i Y_{2,i} I(\boldsymbol{W}_i = \boldsymbol{w}_k)$, $Z_{2,k} = \sum_{i=1}^{n} (1 - T_i) Y_{2,i} I(\boldsymbol{W}_i = \boldsymbol{w}_k)$, $n_{1,k} = \sum_{i=1}^{n} T_i I(\boldsymbol{W}_i = \boldsymbol{w}_k)$, $n_{0,k} = \sum_{i=1}^{n} (1 - T_i) I(\boldsymbol{W}_i = \boldsymbol{w}_k)$, $n_k = \sum_{i=1}^{n} I(\boldsymbol{W}_i = \boldsymbol{w}_k)$, with $I()$ the Indicator function. This would lead to 
\begin{equation*}
    \hat{\beta}_1^* = \log\left(\frac{\sum_{k=1}^{K} \frac{n_{0,k} X_{1,k}}{n_k}}{\sum_{k=1}^{K} \frac{n_{1,k} Z_{1,k}}{n_k}}\right) \quad \text{and} \quad \hat{\beta}_2^* = \log\left(\frac{\sum_{k=1}^{K} \frac{n_{0,k} X_{2,k}}{n_k}}{\sum_{k=1}^{K} \frac{n_{1,k} Z_{2,k}}{n_k}}\right),
\end{equation*}
whose variances and covariance can be obtained from the sandwich formula
\small{\begin{eqnarray*}
\frac{1}{K} \left\{ \frac{1}{K} \sum_{k=1}^{K} \omega_k \frac{\partial \boldsymbol{U}(X_{1,k}, Z_{1,k}, X_{2,k}, Z_{2,k};\boldsymbol{\theta})}{\partial \boldsymbol{\theta}}_{|\boldsymbol{\theta} = \widehat{\boldsymbol{\theta}}} \right\}^{-1}  \left[ \frac{1}{K-1} \sum_{k=1}^{K} \left\{\omega_k \boldsymbol{U}(X_{1,k}, Z_{1,k}, X_{2,k}, Z_{2,k};\widehat{\boldsymbol{\theta}} ) \right\} \right. \\
 \left\{ \omega_k \boldsymbol{U}(X_{1,k}, Z_{1,k}, X_{2,k}, Z_{2,k};\widehat{\boldsymbol{\theta}} ) \right\}^\top \Bigg] \left[ \left\{ \frac{1}{K} \sum_{k=1}^{K} \omega_k \frac{\partial \boldsymbol{U}(X_{1,k}, Z_{1,k}, X_{2,k}, Z_{2,k};\boldsymbol{\theta})}{\partial \boldsymbol{\theta}}_{|\boldsymbol{\theta} = \widehat{\boldsymbol{\theta}}} \right\}^{-1} \right]^\top ,
\end{eqnarray*}}
\normalsize with
$$\frac{\partial \boldsymbol{U}(X_{1,k}, Z_{1,k}, X_{2,k}, Z_{2,k};\boldsymbol{\theta})}{\partial \boldsymbol{\theta}} = \begin{pmatrix}
- \exp(\beta_1^*) \frac{Z_{1,k}}{n_{0,k}}& 0\\
0 & - \exp(\beta_2^*) \frac{Z_{2,k}}{n_{0,k}}
\end{pmatrix}.$$
As a result, if Assumption A$_1$ and Assumption A$_{2,\boldsymbol{W}}$ hold under the setting of Figure \ref{fig:DAGmediation} {(B)} in the Main Document, $\hat \beta_1^* - \hat \beta_2^*$ is an unbiased estimate of the log-natural direct effect. Again, the variance estimate can be obtained from adding the variances of $\hat \beta_1^*$ and $\hat \beta_2^*$ and subtracting two times their covariance.\\

Finally, if $\boldsymbol{W}$ only has a few large strata, Etievant et al. \cite{etievantsampson2023} suggests applying the method in Section \ref{subsec:nontargeted} in the Main Document and detailed in Supplementary Material \ref{appendix:estimationinference} separately in each stratum, and then combining the stratum-specific estimates with weights inversely proportional to their variance estimates (the stratum-specific joint approach, SS-Joint). This requires less stringent assumptions as one would only need models in Equations (\ref{eq:equation1}) and (\ref{eq:equation2}) in Section \ref{subsec:notation} in the Main Document to hold within each stratum of $\boldsymbol{W}$ (with a common $\beta_1$) and Assumption A$_1$ to hold within each stratum of $\boldsymbol{W}$ (with non-zero stratum-specific proportionality constant $k_{\boldsymbol{W}}$).

\newpage

\begin{landscape}

\section{Simulation under the observational setting}

\subsection{Parameter values for the simulation in Section \ref{sec:simul} in the Main Document}\label{appendix:paramval}

\begin{Supplementary Table}[ht]
\centering
\begin{tabular}{ccccccccccccccc}
  \hline
  \hline
& \multicolumn{13}{c}{$w_{\text{age}}$}  \\
& & 15 & 15.5 & 16 & 16.5 & 17 & 17.5 & 18 & 18.5 & 19 & 19.5 & 20 & 20.5 & 21 \\ 
  \hline
\multirow{3}{*}{\rotatebox[origin=c]{90}{$w_{\text{site}}$}} & \rotatebox[origin=c]{90}{0} & 0.076 & 0.076 & 0.081 & 0.076 & 0.076 & 0.081 & 0.081 & 0.076 & 0.076 & 0.076 & 0.081 & 0.070 & 0.076 \\ 
& \rotatebox[origin=c]{90}{1} & 0.077 & 0.083 & 0.077 & 0.077 & 0.077 & 0.077 & 0.071 & 0.077 & 0.077 & 0.077 & 0.071 & 0.077 & 0.077 \\ 
& \rotatebox[origin=c]{90}{2} & 0.082 & 0.076 & 0.076 & 0.076 & 0.076 & 0.076 & 0.076 & 0.082 & 0.076 & 0.076 & 0.076 & 0.076 & 0.076 \\ 
   \hline
   \hline
\end{tabular}
\caption{Parameter values for $\PP(W_{\text{age}} = w_{\text{age}} \mid W_{\text{site}} = w_{\text{site}})$.}\label{webtable:Wage}
\end{Supplementary Table}

\begin{Supplementary Table}[ht]
\centering
\begin{tabular}{ccccccccccccccc}
  \hline
  \hline
& \multicolumn{13}{c}{$w_{\text{age}}$}  \\
& & 15 & 15.5 & 16 & 16.5 & 17 & 17.5 & 18 & 18.5 & 19 & 19.5 & 20 & 20.5 & 21 \\ 
  \hline
\multirow{3}{*}{\rotatebox[origin=c]{90}{$w_{\text{site}}$}} & \rotatebox[origin=c]{90}{0} & 0.088 & 0.120 & 0.106 & 0.072 & 0.161 & 0.193 & 0.228 & 0.224 & 0.273 & 0.248 & 0.221 & 0.252 & 0.233 \\ 
& \rotatebox[origin=c]{90}{1} & 0.092 & 0.085 & 0.137 & 0.169 & 0.212 & 0.253 & 0.189 & 0.252 & 0.296 & 0.252 & 0.270 & 0.319 & 0.266 \\ 
& \rotatebox[origin=c]{90}{2} & 0.121 & 0.141 & 0.211 & 0.216 & 0.214 & 0.208 & 0.226 & 0.268 & 0.243 & 0.220 & 0.294 & 0.275 & 0.340 \\
   \hline
    \hline
\end{tabular}
\caption{Parameter values for $\PP(A = a_{\text{high}} \mid W_{\text{site}} = w_{\text{site}}, W_{\text{age}} = w_{\text{age}})$.}\label{webtable:Ahigh}
\end{Supplementary Table}

\begin{Supplementary Table}[ht]
\centering
\begin{tabular}{ccccccccccccccc}
  \hline
  \hline
& \multicolumn{13}{c}{$w_{\text{age}}$}  \\
& & 15 & 15.5 & 16 & 16.5 & 17 & 17.5 & 18 & 18.5 & 19 & 19.5 & 20 & 20.5 & 21 \\ 
  \hline
\multirow{3}{*}{\rotatebox[origin=c]{90}{$w_{\text{site}}$}} & \rotatebox[origin=c]{90}{0} & 0.292 & 0.312 & 0.300 & 0.476 & 0.394 & 0.415 & 0.439 & 0.390 & 0.489 & 0.440 & 0.527 & 0.612 & 0.533 \\ 
& \rotatebox[origin=c]{90}{1} & 0.342 & 0.368 & 0.325 & 0.381 & 0.388 & 0.372 & 0.528 & 0.452 & 0.456 & 0.458 & 0.535 & 0.507 & 0.601 \\ 
& \rotatebox[origin=c]{90}{2} & 0.260 & 0.226 & 0.264 & 0.319 & 0.406 & 0.397 & 0.467 & 0.428 & 0.451 & 0.453 & 0.479 & 0.495 & 0.471 \\ 
   \hline
    \hline
\end{tabular}
\caption{Parameter values for $\PP(A = a_{\text{medium}} \mid W_{\text{site}} = w_{\text{site}}, W_{\text{age}} = w_{\text{age}})$.}\label{webtable:Amedium}
\end{Supplementary Table}

\begin{Supplementary Table}[ht]
\centering
\begin{tabular}{ccccccccccccccc}
  \hline
  \hline
& \multicolumn{13}{c}{$w_{\text{age}}$}  \\
& & 15 & 15.5 & 16 & 16.5 & 17 & 17.5 & 18 & 18.5 & 19 & 19.5 & 20 & 20.5 & 21 \\ 
  \hline
\multirow{3}{*}{\rotatebox[origin=c]{90}{$w_{\text{site}}$}} & 0 & 0.621 & 0.569 & 0.594 & 0.452 & 0.445 & 0.392 & 0.333 & 0.386 & 0.238 & 0.312 & 0.251 & 0.136 & 0.234 \\ 
& \rotatebox[origin=c]{90}{1} & 0.567 & 0.547 & 0.538 & 0.450 & 0.400 & 0.375 & 0.283 & 0.296 & 0.248 & 0.291 & 0.196 & 0.174 & 0.133 \\ 
& \rotatebox[origin=c]{90}{2} & 0.619 & 0.633 & 0.525 & 0.465 & 0.380 & 0.395 & 0.307 & 0.303 & 0.305 & 0.327 & 0.227 & 0.230 & 0.189 \\
   \hline
    \hline
\end{tabular}
\caption{Parameter values for $\PP(A = a_{\text{low}} \mid W_{\text{site}} = w_{\text{site}}, W_{\text{age}} = w_{\text{age}})$.}\label{webtable:Alow}
\end{Supplementary Table}

\begin{Supplementary Table}[ht]
\centering
\begin{tabular}{ccccc}
  \hline
  \hline
$\alpha_T$ & $\delta_T$ & $\gamma_T$ & $\zeta_T$\\
  \hline
-0.91 & - $\frac{1}{18}$ & 1.5 & 1 \\
   \hline
    \hline
\end{tabular}
\caption{Parameter values used for the simulation of $T$.}\label{webtable:T}
\end{Supplementary Table}

\begin{Supplementary Table}[ht]
\centering
\begin{tabular}{ccccccccccccccc}
  \hline
  \hline
& & \multicolumn{13}{c}{$w_{\text{age}}$}  \\
& & 15 & 15.5 & 16 & 16.5 & 17 & 17.5 & 18 & 18.5 & 19 & 19.5 & 20 & 20.5 & 21 \\ 
  \hline
& $A = a_{\text{low}}$ & 0.017 & 0.017 & 0.017 & 0.017 & 0.017 & 0.017 & 0.017 & 0.018 & 0.018 & 0.018 & 0.018 & 0.018 & 0.018 \\ 
& $A = a_{\text{medium}}$ & 0.167 & 0.168 & 0.170 & 0.171 & 0.173 & 0.174 & 0.176 & 0.177 & 0.179 & 0.180 & 0.182 & 0.183 & 0.185 \\ 
& $A = a_{\text{high}}$ & 0.990 & 0.990 & 0.990 & 0.990 & 0.990 & 0.990 & 0.990 & 0.990 & 0.990 & 0.990 & 0.990 & 0.990 & 0.990  \\ 
   \hline
\end{tabular}
\caption{Parameter values for $\PP(\tilde{A} = a_{\text{high}} \mid W_{\text{age}} = w_{\text{age}}, A, T = 1)$.}\label{webtable:Atildahigh}
\end{Supplementary Table}

\begin{Supplementary Table}[ht]
\centering
\begin{tabular}{ccccccccccccccc}
  \hline
  \hline
& & \multicolumn{13}{c}{$w_{\text{age}}$}  \\
& & 15 & 15.5 & 16 & 16.5 & 17 & 17.5 & 18 & 18.5 & 19 & 19.5 & 20 & 20.5 & 21 \\ 
  \hline
& $A = a_{\text{low}}$ & 0.165 & 0.167 & 0.168 & 0.170 & 0.171 & 0.173 & 0.174 & 0.176 & 0.177 & 0.179 & 0.180 & 0.182 & 0.183 \\ 
& $A = a_{\text{medium}}$ & 0.825 & 0.823 & 0.822 & 0.820 & 0.819 & 0.817 & 0.816 & 0.814 & 0.812 & 0.811 & 0.809 & 0.808 & 0.806 \\ 
& $A = a_{\text{high}}$ & 0.010 & 0.010 & 0.010 & 0.010 & 0.010 & 0.010 & 0.010 & 0.010 & 0.010 & 0.010 & 0.010 & 0.010 & 0.010 \\ 
   \hline
\end{tabular}
\caption{Parameter values for $\PP(\tilde{A} = a_{\text{medium}} \mid W_{\text{age}} = w_{\text{age}}, A, T = 1)$.}\label{webtable:Atildamedium}
\end{Supplementary Table}

\begin{Supplementary Table}[ht]
\centering
\begin{tabular}{ccccccccccccccc}
  \hline
  \hline
& & \multicolumn{13}{c}{$w_{\text{age}}$}  \\
& & 15 & 15.5 & 16 & 16.5 & 17 & 17.5 & 18 & 18.5 & 19 & 19.5 & 20 & 20.5 & 21 \\ 
  \hline
& $A = a_{\text{low}}$ & 0.818 & 0.817 & 0.815 & 0.813 & 0.812 & 0.810 & 0.809 & 0.807 & 0.805 & 0.804 & 0.802 & 0.800 & 0.799 \\ 
& $A = a_{\text{medium}}$ & 0.008 & 0.008 & 0.008 & 0.009 & 0.009 & 0.009 & 0.009 & 0.009 & 0.009 & 0.009 & 0.009 & 0.009 & 0.009 \\ 
& $A = a_{\text{high}}$ & 0.000 & 0.000 & 0.000 & 0.000 & 0.000 & 0.000 & 0.000 & 0.000 & 0.000 & 0.000 & 0.000 & 0.000 & 0.000 \\ 
   \hline
\end{tabular}
\caption{Parameter values for $\PP(\tilde{A} = a_{\text{low}} \mid W_{\text{age}} = w_{\text{age}}, A, T = 1)$.}\label{webtable:Atildalow}
\end{Supplementary Table}

\begin{Supplementary Table}[ht]
\centering
\begin{tabular}{ccccc}
  \hline
  \hline
$\beta_1$ & $\lambda_1$ & $\mu_{W_{\text{site}} = 0}$ & $\mu_{W_{\text{site}} = 1}$ & $\mu_{W_{\text{site}} = 2}$\\
  \hline
 - 0.73 & 0.01 & 0.06 & -0.26 & 0.50\\
    \hline
    \hline
\end{tabular}
\caption{Parameter values used for the simulation of $Y_1$. As a reminder, $\alpha_1$ varies depending on the value of $\PP(Y_1 = 1)$ considered in the scenario.}\label{webtable:Y1}
\end{Supplementary Table}

\end{landscape}

\newpage

\begin{Supplementary Table}[h!]
\centering
\begin{tabular}{cccc}
  \hline
  \hline
$j$ & $\PP(Y_2^{(j)} = 1)$ & $\lambda_2^{(j)}$ & $\mu_2^{(j)}$ \\ 
  \hline
1 & 0.07 & 0.0035 & -0.2504 \\ 
  2 & 0.03 & 0.0026 & -0.1048 \\ 
  3 & 0.0145 & 0.0071 & -0.0994 \\ 
  4 & 0.055 & 0.0156 & -0.3612 \\ 
  5 & 0.075 & 0.0004 & -0.1164 \\ 
  6 & 0.04 & 0.0090 & -0.2218 \\ 
  7 & 0.02 & 0.0073 & -0.2030 \\ 
  8 & 0.055 & 0.0078 & -0.0325 \\ 
  9 & 0.065 & 0.0054 & 0.1126 \\ 
  10 & 0.075 & 0.0015 & 0.3296 \\ 
  11 & 0.09 & 0.0082 & -0.1547 \\ 
  12 & 0.03 & 0.0036 & 0.3212 \\ 
  13 & 0.095 & 0.0085 & -0.2316 \\ 
  14 & 0.02 & 0.0052 & 0.1313 \\ 
  15 & 0.09 & 0.0077 & 0.5098 \\ 
  16 & 0.07 & 0.0120 & -0.0070 \\ 
  17 & 0.04 & 0.0112 & -0.1339 \\ 
  18 & 0.07 & 0.0143 & -0.0015 \\ 
  19 & 0.085 & 0.0011 & 0.3554 \\ 
  20 & 0.06 & 0.0019 & -0.2277 \\ 
   \hline
   \hline
\end{tabular}
\caption{Parameter values used for the simulation of the $Y_2^{(j)}$, $j \in \{1,\dots,N_{NT}\}$. As a reminder, $\alpha_2^{(j)}$ varies depending on the value of $\PP(Y_2^{(j)} = 1)$.}\label{webtable:Y2}
\end{Supplementary Table}

\subsection{Additional results from the simulation in Section \ref{sec:simul} in the Main Document}\label{appendix:simuladditionalres}

Supplementary Table \ref{webtable:confn5000} gives the results of the simulation in Section \ref{sec:simul} in the Main Document for the scenarios with $n = 5, 000$. The results for the scenarios with $n = 10, 000$ are given in Table \ref{table:conf} in Section \ref{sec:simul} in the Main Document.

In Table \ref{table:conf} and Supplementary Table \ref{webtable:confn5000}, the relative bias is adequately computed with respect to the true log-natural direct effect of the vaccine on the targeted outcome. To better illustrate that the true average total effect and natural direct effect are relatively different, Supplementary Table \ref{webtable:biasATE} also displays for each scenario the mean relative bias with respect to the log-average total effect over the 5,000 studies. When computed with respect to the average total effect, the relative bias of methods Joint-MH and Joint-Reg is between 6 and 16 times higher than when computed with respect to the natural direct effect. In addition, consider for example the scenario with $\PP(Y_1 = 1) = 0.14$ and ($a_{\text{low}}, a_{\text{medium}}, a_{\text{high}}$) = (0,1,2.5); the vaccine efficacy derived from the true natural direct effect is $\approx 0.52$, whereas that derived from the average total effect is $\approx 0.42$. Such a difference in vaccine efficacy in a non-inferiority trial could lead to changes in vaccine recommendations. In particular, note that the difference in vaccine efficacy is larger than those found for incident HPV 16 and 18 infections in a study comparing single-dose, two-dose, and three-dose vaccination schedules \cite{basu2021observational}.

\begin{landscape}
\begin{Supplementary Table}[ht]
\centering
\begin{tabular}{ccccccccccc}
\hline
\hline
\multicolumn{4}{c}{\multirow{2}{*}{Relative bias}}  & \multicolumn{2}{c}{Sample standard} & \multicolumn{2}{c}{Mean sandwich} &  \\
\multirow{2}{*}{Joint-MH} & \multirow{2}{*}{Joint-Reg} & \multirow{2}{*}{MH} & \multirow{2}{*}{Reg} & \multicolumn{2}{c}{deviation} & \multicolumn{2}{c}{standard error} & P($Y_1 = 1$) & ($a_{\text{low}}, a_{\text{medium}}, a_{\text{high}}$) & corr($Y_1,Y_2$) \\ 
 & & & & Joint-MH & Joint-Reg & Joint-MH & Joint-Reg &  \\ 
  \hline
0.022 & 0.024 & 1.223 & 1.217 & 0.084 & 0.082 & 0.084 & 0.083 & 0.14 & (0,1,2.5) & 0.250 \\ 
  0.022 & 0.030 & 0.970 & 0.968 & 0.084 & 0.082 & 0.084 & 0.082 & 0.14 & (0,1,2) & 0.224 \\ 
  0.016 & 0.023 & 0.781 & 0.783 & 0.084 & 0.082 & 0.083 & 0.081 & 0.14 & (0,0.75,1.5) & 0.226 \\ 
  0.024 & 0.030 & 1.225 & 1.223 & 0.148 & 0.144 & 0.148 & 0.146 & 0.05 & (0,1,2.5) & 0.142 \\ 
  0.020 & 0.032 & 0.967 & 0.970 & 0.148 & 0.143 & 0.147 & 0.144 & 0.05 & (0,1,2) & 0.127 \\ 
  0.017 & 0.027 & 0.782 & 0.787 & 0.148 & 0.144 & 0.146 & 0.144 & 0.05 & (0,0.75,1.5) & 0.128 \\ 
  0.027 & 0.034 & 1.229 & 1.227 & 0.214 & 0.208 & 0.213 & 0.209 & 0.025 & (0,1,2.5) & 0.099 \\ 
  0.020 & 0.032 & 0.967 & 0.971 & 0.216 & 0.209 & 0.211 & 0.206 & 0.025 & (0,1,2) & 0.089 \\ 
  0.016 & 0.028 & 0.781 & 0.787 & 0.212 & 0.206 & 0.210 & 0.207 & 0.025 & (0,0.75,1.5) & 0.090 \\ 
   \hline
   \hline
\end{tabular}
\caption{Mean relative bias, sample standard deviation, and mean sandwich standard error, for the estimates proposed by Etievant et al. \cite{etievantsampson2023} over 5,000 simulated studies under the observational setting of Figure \ref{fig:DAGmediation} {(B)} in the Main Document and presented in Section \ref{sec:simul} in the Main Document, with $n = $ 5,000. Relative bias is computed with respect to the true log-natural direct effect of the vaccine on the targeted outcome. Methods Joint-MH and Joint-Reg use targeted and non-targeted infections. As a comparison, the mean relative bias with methods MH and Reg that do not use non-targeted infections is also shown.}\label{webtable:confn5000}
\end{Supplementary Table}
\end{landscape}

\newpage

\begin{landscape}
\begin{Supplementary Table}[ht]
\centering
\begin{tabular}{cccccccc}
  \hline
    \hline
\multicolumn{4}{c}{Relative bias with respect to} \\
\multicolumn{2}{c}{log-NDE} &  \multicolumn{2}{c}{log-ATE} & $n$ & P($Y_1 = 1$) & ($a_{\text{low}}, a_{\text{medium}}, a_{\text{high}}$) & corr($Y_1,Y_2$) \\
Joint-MH & Joint-Reg & Joint-MH & Joint-Reg  \\ 
  \hline
0.022 & 0.024 & 0.290 & 0.287 & 5000 & 0.14 & (0,1,2.5) & 0.250 \\ 
  0.024 & 0.026 & 0.287 & 0.285 & 10000 & 0.14 & (0,1,2.5) & 0.250 \\ 
  0.022 & 0.030 & 0.248 & 0.239 & 5000 & 0.14 & (0,1,2) & 0.224 \\ 
  0.023 & 0.030 & 0.247 & 0.239 & 10000 & 0.14 & (0,1,2) & 0.224 \\ 
  0.016 & 0.023 & 0.256 & 0.247 & 5000 & 0.14 & (0,0.75,1.5) & 0.226 \\ 
  0.017 & 0.024 & 0.255 & 0.246 & 10000 & 0.14 & (0,0.75,1.5) & 0.226 \\ 
  0.024 & 0.030 & 0.288 & 0.279 & 5000 & 0.05 & (0,1,2.5) & 0.142 \\ 
  0.028 & 0.034 & 0.282 & 0.274 & 10000 & 0.05 & (0,1,2.5) & 0.142 \\ 
   \hline
     \hline
\end{tabular}
\caption{Mean relative bias with respect to the true log-natural direct effect and mean relative bias with respect to the true log-average total effect, for the estimates proposed by Etievant et al. \cite{etievantsampson2023} over 5,000 simulated studies under the observational setting of Figure \ref{fig:DAGmediation} {(B)} in the Main Document and presented in Section \ref{sec:simul} in the Main Document. Abbreviations: ATE, average total effect; NDE, natural direct effect.}\label{webtable:biasATE}
\end{Supplementary Table}
\end{landscape}

\newpage

\begin{landscape}

\begin{Supplementary Table}[ht]
\centering
\caption*{Supplementary Table \ref{webtable:biasATE} continued}
\begin{tabular}{cccccccc}
  \hline
    \hline
\multicolumn{4}{c}{Relative bias with respect to} \\
\multicolumn{2}{c}{log-NDE} &  \multicolumn{2}{c}{log-ATE} & $n$ & P($Y_1 = 1$) & ($a_{\text{low}}, a_{\text{medium}}, a_{\text{high}}$) & corr($Y_1,Y_2$) \\
Joint-MH & Joint-Reg & Joint-MH & Joint-Reg  \\ 
  \hline
    0.020 & 0.032 & 0.251 & 0.236 & 5000 & 0.05 & (0,1,2) & 0.127 \\ 
  0.025 & 0.035 & 0.245 & 0.232 & 10000 & 0.05 & (0,1,2) & 0.127 \\ 
  0.017 & 0.027 & 0.255 & 0.242 & 5000 & 0.05 & (0,0.75,1.5) & 0.128 \\ 
  0.015 & 0.024 & 0.258 & 0.246 & 10000 & 0.05 & (0,0.75,1.5) & 0.128 \\ 
  0.027 & 0.034 & 0.283 & 0.274 & 5000 & 0.025 & (0,1,2.5) & 0.099 \\ 
  0.029 & 0.036 & 0.281 & 0.271 & 10000 & 0.025 & (0,1,2.5) & 0.099 \\ 
  0.020 & 0.032 & 0.251 & 0.236 & 5000 & 0.025 & (0,1,2) & 0.089 \\ 
  0.026 & 0.037 & 0.243 & 0.229 & 10000 & 0.025 & (0,1,2) & 0.089 \\ 
  0.016 & 0.028 & 0.256 & 0.241 & 5000 & 0.025 & (0,0.75,1.5) & 0.090 \\ 
  0.014 & 0.024 & 0.258 & 0.246 & 10000 & 0.025 & (0,0.75,1.5) & 0.090 \\ 
   \hline
     \hline
\end{tabular}
\end{Supplementary Table}
\end{landscape}

\newpage

\subsection{Additional simulation}\label{webapp:additionalsimul}

A similar simulation as in Section \ref{sec:simul} in the Main Document is run, but with parameter values as in Supplementary Table \ref{webtable:T2} below instead of Supplementary Table \ref{webtable:T} in Supplementary Material \ref{appendix:paramval}.

\begin{Supplementary Table}[ht]
\centering
\begin{tabular}{ccccc}
  \hline
  \hline
$\alpha_T$ & $\delta_T$ & $\gamma_T$ & $\zeta_T$\\
  \hline
-0.91 & - $\frac{1}{18}$ & 1.5 & -1 \\
   \hline
    \hline
\end{tabular}
\caption{Parameter values used for the simulation of $T$.}\label{webtable:T2}
\end{Supplementary Table}

The results for the scenarios with $n = 5, 000$ and $n = 10, 000$ are given in Supplementary Tables \ref{webtable:conf2n5000} and \ref{webtable:conf2n10000}, respectively. The boxplots of the estimates obtained with the different methods are displayed in Supplementary Figure \ref{webfig:conf2}. Again, methods Joint-MH and Joint-Reg allow estimation of the log-natural direct effect of the vaccine on the targeted outcome with small bias (columns 1 and 2), even though Assumption A$_2$ is not met. In addition, the estimates obtained with methods Joint-MH or Joint-Reg are closer to $\log\{NDE(Y_1, T=0)\}$ than to $\log\{ATE(Y_1)\}$. As a remark, in comparison to the simulation setting in Section \ref{sec:simul}, the confounding bias goes in the other direction and leads to an overestimated protective effect of the vaccine, because women engaging in riskier sexual behavior are less likely to get the vaccine ($\zeta_T = -1$, see Supplementary Table \ref{webtable:T2}).
 
\begin{landscape}
\begin{Supplementary Table}[ht]
\centering
\begin{tabular}{ccccccccccc}
\hline
\hline
\multicolumn{4}{c}{\multirow{2}{*}{Relative bias}}  & \multicolumn{2}{c}{Sample standard} & \multicolumn{2}{c}{Mean sandwich} &  \\
\multirow{2}{*}{Joint-MH} & \multirow{2}{*}{Joint-Reg} & \multirow{2}{*}{MH} & \multirow{2}{*}{Reg} & \multicolumn{2}{c}{deviation} & \multicolumn{2}{c}{standard error} & P($Y_1 = 1$) & ($a_{\text{low}}, a_{\text{medium}}, a_{\text{high}}$) & corr($Y_1,Y_2$) \\ 
 & & & & Joint-MH & Joint-Reg & Joint-MH & Joint-Reg &  \\ 
  \hline
0.023 & 0.024 & 0.666 & 0.671 & 0.110 & 0.107 & 0.108 & 0.107 & 0.14 & (0,1,2.5) & 0.285 \\ 
  0.006 & 0.005 & 0.486 & 0.486 & 0.104 & 0.102 & 0.103 & 0.102 & 0.14 & (0,1,2) & 0.244 \\ 
  0.012 & 0.012 & 0.326 & 0.326 & 0.098 & 0.095 & 0.096 & 0.095 & 0.14 & (0,0.75,1.5) & 0.242 \\ 
  0.032 & 0.033 & 0.675 & 0.680 & 0.191 & 0.187 & 0.188 & 0.189 & 0.05 & (0,1,2.5) & 0.162 \\ 
  0.019 & 0.020 & 0.499 & 0.501 & 0.186 & 0.182 & 0.180 & 0.180 & 0.05 & (0,1,2) & 0.139 \\ 
  0.023 & 0.023 & 0.337 & 0.338 & 0.171 & 0.167 & 0.169 & 0.168 & 0.05 & (0,0.75,1.5) & 0.138 \\ 
  0.047 & 0.048 & 0.691 & 0.695 & 0.279 & 0.275 & 0.271 & 0.273 & 0.025 & (0,1,2.5) & 0.113 \\ 
  0.037 & 0.035 & 0.517 & 0.516 & 0.269 & 0.264 & 0.259 & 0.259 & 0.025 & (0,1,2) & 0.097 \\ 
  0.037 & 0.035 & 0.351 & 0.349 & 0.242 & 0.237 & 0.242 & 0.242 & 0.025 & (0,0.75,1.5) & 0.096 \\ 
   \hline
   \hline
\end{tabular}
\caption{Mean relative bias, sample standard deviation, and mean sandwich standard error, for the estimates proposed by Etievant et al. \cite{etievantsampson2023} over 5,000 simulated studies under the observational setting of Figure \ref{fig:DAGmediation} {(B)} in the Main Document and presented in Supplementary Material \ref{webapp:additionalsimul}, with $n = $ 5,000. Relative bias is computed with respect to the true log-natural direct effect of the vaccine on the targeted outcome. Methods Joint-MH and Joint-Reg use targeted and non-targeted infections. As a comparison, the mean relative bias with methods MH and Reg that do not use non-targeted infections is also shown.}\label{webtable:conf2n5000}
\end{Supplementary Table}
\end{landscape}

\begin{landscape}
\begin{Supplementary Table}[ht]
\centering
\begin{tabular}{ccccccccccc}
\hline
\hline
\multicolumn{4}{c}{\multirow{2}{*}{Relative bias}}  & \multicolumn{2}{c}{Sample standard} & \multicolumn{2}{c}{Mean sandwich} &  \\
\multirow{2}{*}{Joint-MH} & \multirow{2}{*}{Joint-Reg} & \multirow{2}{*}{MH} & \multirow{2}{*}{Reg} & \multicolumn{2}{c}{deviation} & \multicolumn{2}{c}{standard error} & P($Y_1 = 1$) & ($a_{\text{low}}, a_{\text{medium}}, a_{\text{high}}$) & corr($Y_1,Y_2$) \\ 
 & & & & Joint-MH & Joint-Reg & Joint-MH & Joint-Reg &  \\ 
  \hline
0.018 & 0.020 & 0.663 & 0.669 & 0.079 & 0.077 & 0.077 & 0.076 & 0.14 & (0,1,2.5) & 0.285 \\ 
  0.007 & 0.007 & 0.488 & 0.488 & 0.074 & 0.073 & 0.073 & 0.073 & 0.14 & (0,1,2) & 0.244 \\ 
  0.008 & 0.008 & 0.323 & 0.324 & 0.070 & 0.068 & 0.069 & 0.068 & 0.14 & (0,0.75,1.5) & 0.242 \\ 
  0.023 & 0.026 & 0.667 & 0.674 & 0.138 & 0.136 & 0.134 & 0.134 & 0.05 & (0,1,2.5) & 0.162 \\ 
  0.014 & 0.015 & 0.495 & 0.497 & 0.131 & 0.128 & 0.128 & 0.128 & 0.05 & (0,1,2) & 0.139 \\ 
  0.012 & 0.014 & 0.328 & 0.330 & 0.122 & 0.120 & 0.120 & 0.119 & 0.05 & (0,0.75,1.5) & 0.138 \\ 
  0.031 & 0.034 & 0.676 & 0.682 & 0.196 & 0.193 & 0.193 & 0.193 & 0.025 & (0,1,2.5) & 0.113 \\ 
  0.020 & 0.021 & 0.501 & 0.502 & 0.187 & 0.183 & 0.184 & 0.184 & 0.025 & (0,1,2) & 0.097 \\ 
  0.016 & 0.017 & 0.332 & 0.333 & 0.176 & 0.173 & 0.172 & 0.171 & 0.025 & (0,0.75,1.5) & 0.096 \\ 
   \hline
   \hline
\end{tabular}
\caption{Mean relative bias, sample standard deviation, and mean sandwich standard error, for the estimates proposed by Etievant et al. \cite{etievantsampson2023} over 5,000 simulated studies under the observational setting of Figure \ref{fig:DAGmediation} {(B)} in the Main Document and presented in Supplementary Material \ref{webapp:additionalsimul}, with $n = $ 10,000. Relative bias is computed with respect to the true log-natural direct effect of the vaccine on the targeted outcome. Methods Joint-MH and Joint-Reg use targeted and non-targeted infections. As a comparison, the mean relative bias with methods MH and Reg that do not use non-targeted infections is also shown.}\label{webtable:conf2n10000}
\end{Supplementary Table}
\end{landscape}

\newpage

\begin{Supplementary Figure}[h!]
    \centering
    \includegraphics[width=1\linewidth]{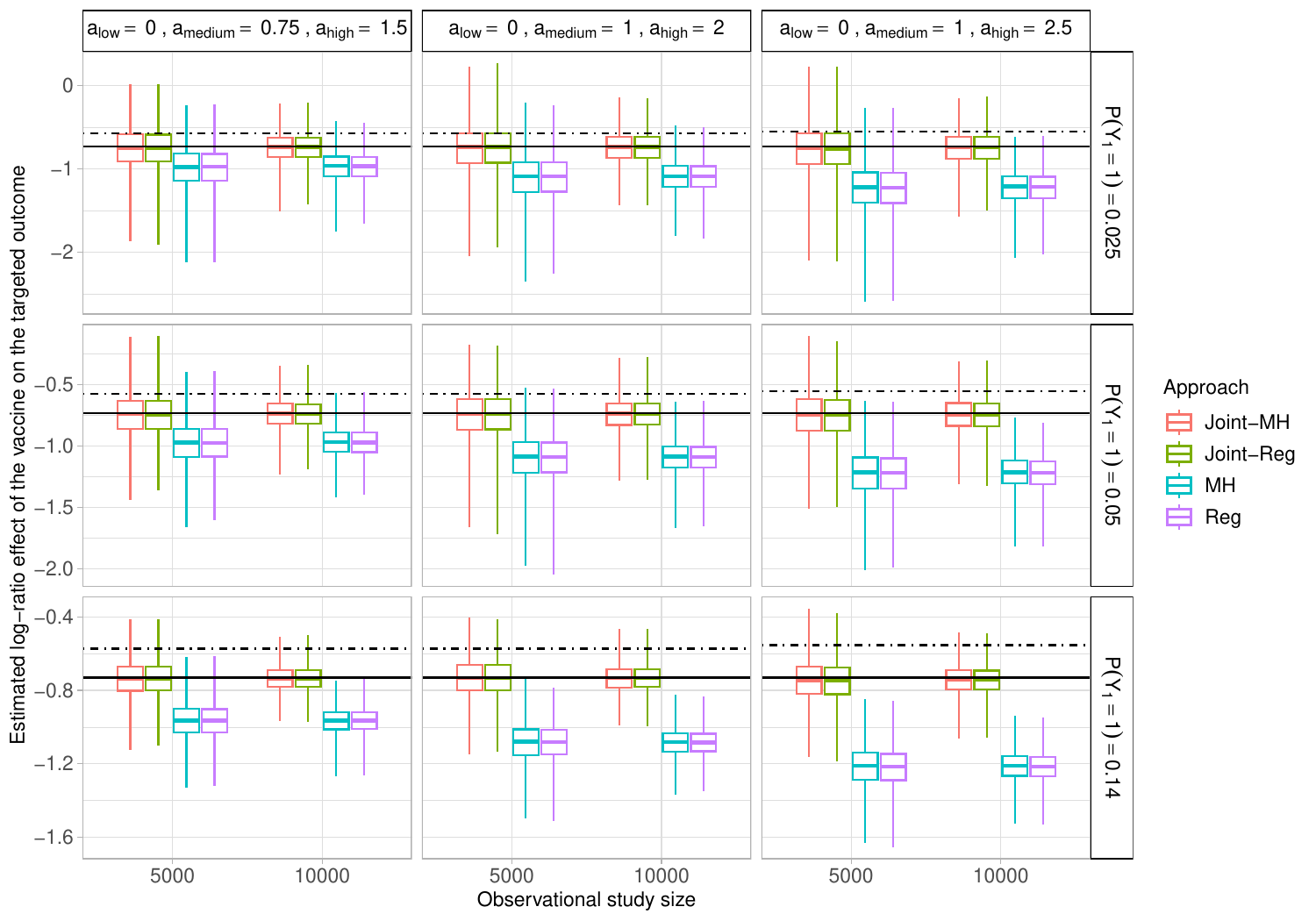}
    \caption{Boxplots of estimates with Joint-MH (red), Joint-Reg (green), MH (blue), and Reg (purple), from 5,000 simulated studies under the observational setting of Figure \ref{fig:DAGmediation} {(B)} in the Main Document and presented in Supplementary Material \ref{webapp:additionalsimul} and in various scenarios. The true log-natural direct effect of $T$ on $Y_1$ is indicated by the horizontal black line, while the true log-average total effect is indicated by the dash-dotted line.} \label{webfig:conf2}
\end{Supplementary Figure}

\newpage

\section{Connection with estimates from a randomized con-trolled trial}\label{sec:connection}

The natural direct effect of the vaccine can be estimated, provided data on infections with human papillomavirus (HPV) nonvaccine targeted strains is available and under certain assumptions; see Section \ref{sec:method} in the Main Document. Such a quantity has a clear causal meaning; it represents the immunological effect of the vaccine on the targeted HPV strain. But is it really relevant from an public health perspective? Changes of behaviors (i.e., risk compensation) can impact the effectiveness of the vaccine, so should not they be accounted for? In particular, if the indirect effect of the vaccine through the change of sexual behavior is negative, the average total effect of the vaccine would show an attenuated protective effect compared to the natural direct effect. In other words, the direct effect of the vaccine would suggest higher protection than what women will effectively experience. Unfortunately, under the settings of Figure \ref{fig:DAGmediation} {(A)} or Figure \ref{fig:DAGmediation} {(B)} in the Main Document, the average total effect of the vaccine on the targeted HPV infections cannot be assessed because of unmeasured confounder $\boldsymbol{A}$. 

Now imagine that a blinded randomized controlled trial (RCT) would have been performed. Because the vaccine would be randomly assigned, confounding bias would not be an issue. But because the participants would not know if the vaccine or the placebo is administered, the vaccine would not have a behavioral impact on the risk of HPV infections. The relationships among the variables under such a blinded RCT are assumed to be as in Supplementary Figure \ref{fig:DAGrandom} {(A)}, where to simplify $\boldsymbol{W}=\emptyset$. Under this setting, the average total effect could be estimated from contrast $\frac{\E(Y_{1} \mid T = 1)}{\E(Y_{1} \mid T = 0)}$, i.e., from data on $\{T,Y_1\}$ only (see Supplementary Material \ref{appendix:identifiabilityRCT} for details), but it would also only quantify the immunological effect of the vaccine. 

\begin{Supplementary Figure}
\begin{minipage}[c]{0.5\linewidth}
\centering
\includegraphics[width=0.8\linewidth]{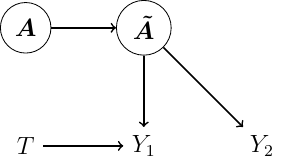}
\end{minipage}\hfill
\begin{minipage}[c]{0.5\linewidth}
\centering
\includegraphics[width=0.8\linewidth]{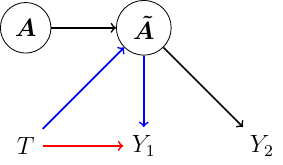}
\end{minipage}

\vspace{0.4cm}

\begin{minipage}[c]{0.5\linewidth}
\centering
{(A)}
\end{minipage}\hfill
\begin{minipage}[c]{0.5\linewidth}
\centering
{(B)}
\end{minipage}

\caption{Graph depicting the relationships among the variables in - {(A)} The blinded randomized controlled trial setting. - {(B)} The unblinded randomized controlled trial setting. Circled variables are unmeasured.}\label{fig:DAGrandom}
\end{Supplementary Figure}

The blinding precludes behavioral changes after vaccination, so an idea would be to perform an unblinded RCT. Again, the vaccine would be randomly assigned so confounding bias would not be an issue. But there could be a behavioral impact of the vaccine, as participants would know if they receive the vaccine or the placebo. The relationships among the variables under such an unblinded RCT are assumed to be as in Supplementary Figure \ref{fig:DAGrandom} {(B)}. Under this setting, the average total effect of $T$ on $Y_1$ captures both the immunological and behavioral effect of the vaccine, just as in the observational setting. But here it could be estimated from $\frac{\E(Y_{1} \mid T = 1)}{\E(Y_{1} \mid T = 0)}$, i.e., from data on $\{T,Y_1\}$ only; see Supplementary Material \ref{appendix:identifiabilityunblindRCT} for details. 

Finally, note that quantifying the behavioral effect of the vaccine can also be of interest, as a large effect would suggest that, e.g., more education about HPV vaccines and HPV infections is needed. Unfortunately, because $\boldsymbol{\tilde{A}}$ is unobserved, under the unblinded RCT setting of Supplementary Figure \ref{fig:DAGrandom} {(B)} one would usually not be able to know the portion of the total effect that is mediated through the change of behavior. But again, under this setting $T$ would also have an effect on $Y_2$ through the change of behavior. Using similar derivations as for $Y_1$, the average total effect of $T$ on $Y_2$ could be estimated from $\frac{\E(Y_{2} \mid T = 1)}{\E(Y_{2} \mid T = 0)}$. On the other hand, under Assumption A$_1$ (given in Section \ref{subsec:nontargeted} in the Main Document), the behavioral (i.e., indirect) effect of the vaccine is the same on the vaccine targeted and non-targeted strains. Therefore, as the effect of $T$ on $Y_2$ is fully mediated by $\boldsymbol{\tilde{A}}$, $\frac{\E(Y_{2} \mid T = 1)}{\E(Y_{2} \mid T = 0)}$ could be used to quantify the natural indirect effect of $T$ on $Y_1$; see Supplementary Material \ref{appendix:identifiabilityunblindRCT} for more details.

\newpage 

\section{Identifiability of the average total effect under the blinded RCT setting}\label{appendix:identifiabilityRCT}

Under the blinded RCT setting of Supplementary Figure \ref{fig:DAGrandom} {(A)}, the following identifiability conditions hold, for any value $t$:
\begin{eqnarray*}
   & \text{\tiny{(Consistency assumption)}} \quad & \text{If } T=t \text{, then }Y_1^{t} = Y_1,\\
   & \text{\tiny{(Ignorability assumption)}} \quad & Y_1^{t} \indep T.
\end{eqnarray*}
In this way, $\E \left( Y_1^{t} \right) = \E\left( Y_1 \mid T=t \right)$ and $ATE(Y_1)$ can be estimated from data on $T$ and $Y_1$ only, with $\frac{\E\left( Y_1 \mid T=1 \right)}{\E\left( Y_1 \mid T= 0\right)}$. 

Then because $T \indep \boldsymbol{\tilde{A}}$ under the setting of Supplementary Figure \ref{fig:DAGrandom} {(A)}, $\E \left( Y_1^{t} \right)= \E_{\boldsymbol{\tilde{A}}} \Big\{ \E\Big( Y_1 \mid T=t, \boldsymbol{\tilde{A}} \Big)\Big\}$, and finally, using Equation (\ref{eq:equation1}) in Section \ref{subsec:notation} in the Main Document, $ATE(Y_1) = \exp(\beta_1)$. Therefore, the average total effect of the vaccine on the targeted HPV strain under the blinded RCT setting of Supplementary Figure \ref{fig:DAGrandom} {(A)} differs from that under the observational setting of Figure \ref{fig:DAGmediation} {(A)} in the Main Document. More precisely, it coincides with the natural direct effect of the vaccine under the observational setting of Figure \ref{fig:DAGmediation} {(A)}; see Equations (\ref{eq:ATE}) and (\ref{eq:NDE}) in Section \ref{subsec:causaleffects} in the Main Document.

\newpage

\section{Identifiability of the average total effects under the unblinded RCT setting}\label{appendix:identifiabilityunblindRCT}

Under the unblinded RCT setting of Supplementary Figure \ref{fig:DAGrandom} {(B)} the same identifiability conditions as under the blinded RCT setting of Supplementary Figure \ref{fig:DAGrandom} {(A)} in the Main Document hold. Thus again, for any $t$, $\E \left( Y_1^{t} \right) = \E\left( Y_1 \mid T=t \right)$, and $ATE(Y_1)$ can be estimated from data on $T$ and $Y_1$ only, with $\frac{\E\left( Y_1 \mid T=1 \right)}{\E\left( Y_1 \mid T= 0\right)}$. 

However here, $T \nindep \boldsymbol{\tilde{A}}$, $T \indep \boldsymbol{{A}}$, and $\boldsymbol{{A}} \indep Y_1 \mid T, \boldsymbol{\tilde{A}}$. Thus $\E \left( Y_1^{t} \right)
    = \E_{\boldsymbol{{A}}} \Big[\E_{\boldsymbol{\tilde{A}}} \Big\{ \E \Big( Y_1 \mid $  $T=t, \boldsymbol{\tilde{A}} \Big)\mid \boldsymbol{{A}}, T= t\Big\}\Big]$ and finally, using Equation (\ref{eq:equation1}) in Section \ref{subsec:notation} in the Main Document, $ATE(Y_1) = \exp(\beta_1) \times \frac{\E_{\boldsymbol{{A}}}\big[\E \big\{ g_1(\boldsymbol{\tilde{A}})\mid \boldsymbol{A}, T=1\big\} \big]}{\E_{\boldsymbol{{A}}}\big[\E \big\{ g_1(\boldsymbol{\tilde{A}})\mid \boldsymbol{A}, T=0\big\} \big]}$. Therefore, the average total effect of the vaccine on the targeted HPV strain under the unblinded RCT setting of Supplementary Figure \ref{fig:DAGrandom} {(B)} coincides with that under the observational setting of Figure \ref{fig:DAGmediation} {(A)} in the Main Document; see Equation (\ref{eq:ATE}) in Section \ref{subsec:causaleffects} in the Main Document. The natural direct effects, and thus the natural indirect effects, coincide too.

In fact, under the unblinded RCT setting of Supplementary Figure \ref{fig:DAGrandom} {(B)}, the same identifiability and independence conditions hold for $Y_2$, so that $ATE(Y_2) \equiv \frac{\E\left( Y_2^{T=1} \right)}{\E\left( Y_2^{T=0}\right)} = \frac{\E\left( Y_2 \mid T=1 \right)}{\E\left( Y_2 \mid T=0 \right)}$, and using Equation (\ref{eq:equation2}) in Section \ref{subsec:notation} in the Main Document, it further equals $\frac{\E_{\boldsymbol{{A}}}\big[\E \big\{ g_2(\boldsymbol{\tilde{A}})\mid \boldsymbol{A}, T=1\big\} \big]}{\E_{\boldsymbol{{A}}}\big[\E \big\{ g_2(\boldsymbol{\tilde{A}})\mid \boldsymbol{A}, T=0\big\} \big]}$. As a result, under the setting of Supplementary Figure \ref{fig:DAGrandom} {(B)} and if Assumption A$_1$ given in Section \ref{subsec:nontargeted} in the Main Document holds, $\frac{\E\left( Y_2 \mid T=1 \right)}{\E\left( Y_2 \mid T= 0\right)}$ coincides with $NIE(Y_1,T=1)$, and $\frac{\E\left( Y_1 \mid T=1 \right)}{\E\left( Y_1 \mid T= 0\right)} \times \frac{\E\left( Y_2 \mid T=0 \right)}{\E\left( Y_2 \mid T= 1\right)}$ coincides with $NDE(Y_1, T=0)$. 

\newpage

\section{Simulation under the blinded randomized controlled trial setting}\label{appendix:simulunblind}

The goal is to verify if, under the unblinded randomized controlled trial (RCT) setting of Supplementary Figure \ref{webfig:DAGrandomW}, using $Y_2$ allows to assess the natural indirect effect of $T$ on $Y_1$ through $\boldsymbol{\tilde{A}}$. A similar simulation setting as in Section \ref{sec:simul} in the Main Document is considered, except that $T$ is randomly assigned. More precisely, $T$ is generated such that $P(T = 1) = \frac{1}{2}$. Assumption A$_1$ given in Section \ref{subsec:nontargeted} in the Main Document also holds.

\begin{Supplementary Figure}[h!]
\centering
\includegraphics[width=0.5\linewidth]{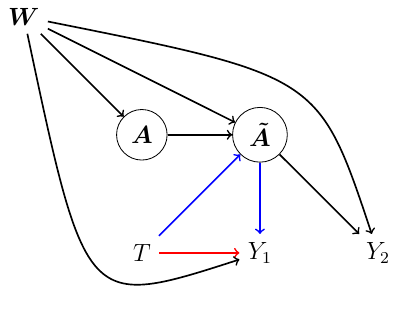}
    \caption{Graph depicting the relationships among the variables in the unblinded randomized controlled trial setting, with measured covariates. Circled variables are unmeasured.}
    \label{webfig:DAGrandomW}
\end{Supplementary Figure}

Method Joint-NC proposed by Etievant et al. \cite{etievantsampson2023} is used to estimate $\beta_1^*$, $\beta_2^*$ and $\beta_1^* - \beta_2^*$, and with standard errors estimated from the sandwich formula; see Supplementary Material \ref{appendix:estimationinference}. As a first remark, $\boldsymbol{W}$ is not a confounder for the $T-Y_1$ or $T-Y_2$ relationships under the unblinded RCT setting of Supplementary Figure \ref{webfig:DAGrandomW}; this is why the simplest Joint method that does not utilize $\boldsymbol{W}$ is used. Then, as a reminder, $\beta_1^* = \log \left\{\frac{\E(Y_{1} \mid T = 1)}{\E(Y_{1} \mid T = 0)} \right\}$ and $\beta_2^* = \log \left\{\frac{\E(Y_{2} \mid T = 1)}{\E(Y_{2} \mid T = 0)} \right\}$; see Supplementary Material \ref{appendix:estimationinference}. Thus, and following Supplementary Material \ref{sec:connection}, the true value of $\log\{ATE(Y_1, T = 0)\}$ is used to compute the mean relative bias for $\hat \beta_1^*$ over the 5,000 studies in each scenario. Similarly, the mean relative bias for $\hat \beta_2^*$ over the 5,000 studies is computed using the true value of $\log\{NIE(Y_1, T = 1)\}$. Finally, as $\log\{ATE(Y_1, T = 0)\} = \log\{NIE(Y_1, T = 1)\ + \log\{NDE(Y_1, T = 0)\}$, the relative bias for $\hat \beta_1^* - \hat \beta_2^*$ is based on the true value of $\log\{NDE(Y_1, T = 0)\}$.

The results for the scenarios with $n = 5, 000$ and $n = 10, 000$ are given in Supplementary Table \ref{webtable:unblindn5000} and Supplementary Table \ref{webtable:unblind}, respectively. Under the unblinded RCT setting of Supplementary Figure \ref{webfig:DAGrandomW}, $ATE(Y_1)$ can be estimated from data on $T$ and $Y_1$ only (column 3), as expected. In addition, $\hat \beta_2^*$ is an estimate of $\log\{ATE(Y_2)\}$, and because Assumption A$_1$ holds, it coincides with $\log\{NIE(Y_1, T = 1)\}$ (column 2). Therefore, from data on $\{T,Y_1, Y_2\}$ one can estimate the log-average total effect of the vaccine on the targeted outcome, but also its log-natural indirect effect and log-natural direct effect (columns 1 to 3). In addition, the sandwich formula allows appropriate estimation of the standard errors (columns 4 to 9). A visual representation of the results for $n = 5, 000$ and $n = 10, 000$ is also provided in Supplementary Figure \ref{webfig:unblind}.

\begin{Supplementary Figure}
    \centering
    \includegraphics[width=1\linewidth]{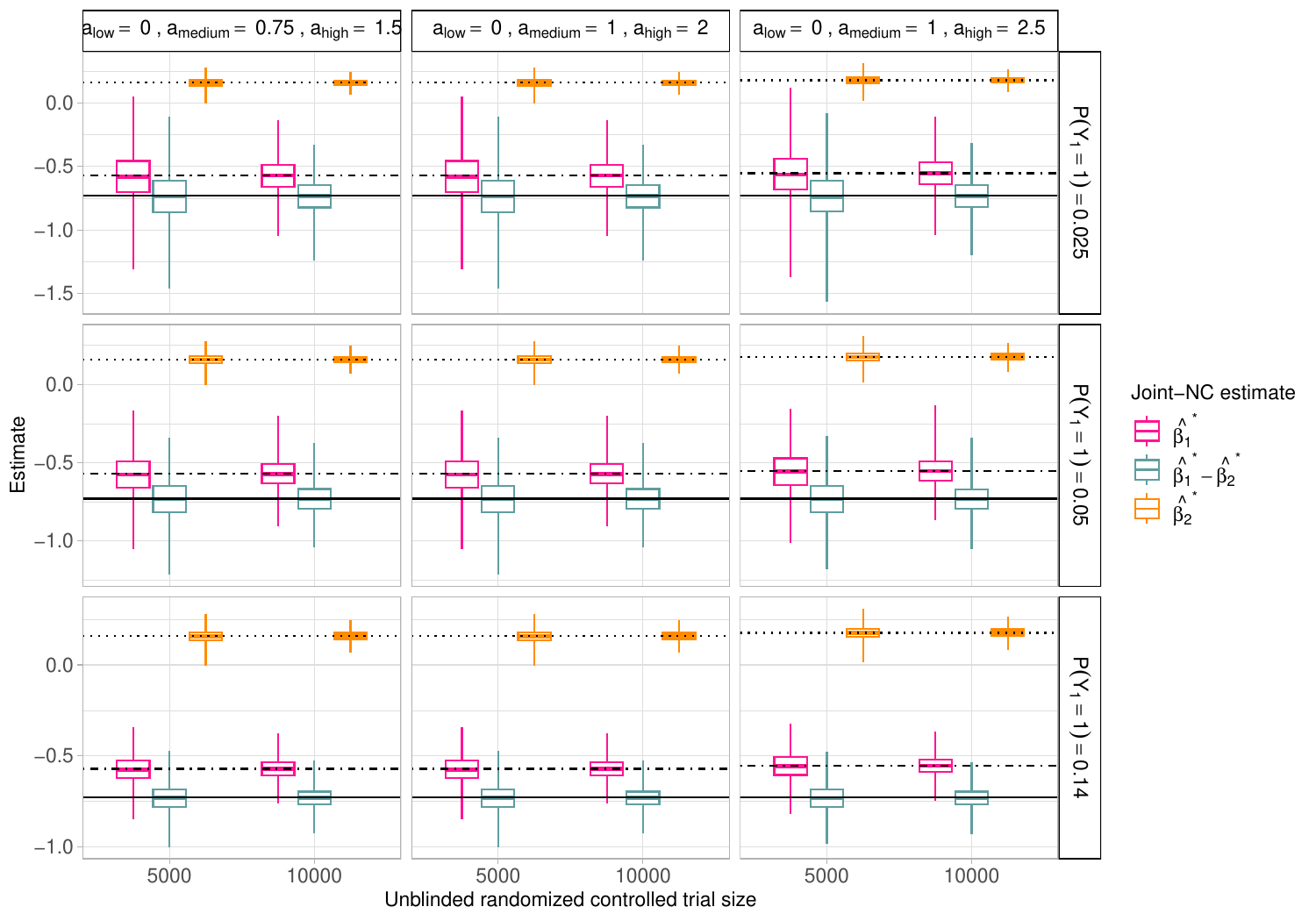}
    \caption{Boxplots of Joint-NC estimates from 5,000 simulated studies under the unblinded randomized controlled trial setting of Supplementary Figure \ref{webfig:DAGrandomW} and in various scenarios. The true log-natural direct effect of $T$ on $Y_1$ is indicated by the horizontal black line, the true log-average total effect of $T$ on $Y_1$ is indicated by the dash-dotted line, and the true log-natural indirect effect of $T$ on $Y_1$ is indicated by the dotted line.}
    \label{webfig:unblind}
\end{Supplementary Figure}

\newpage

\begin{landscape}

\begin{Supplementary Table}[ht]
\centering
\begin{tabular}{cccccccccccc}
\hline
\hline
\multicolumn{3}{c}{\multirow{2}{*}{Relative bias}} & \multicolumn{3}{c}{Sample standard} & \multicolumn{3}{c}{Mean sandwich}  \\
\multicolumn{3}{c}{} & \multicolumn{3}{c}{deviation} & \multicolumn{3}{c}{standard error} & P($Y_1 = 1$) & ($a_{\text{low}}, a_{\text{medium}}, a_{\text{high}}$) & corr($Y_1,Y_2$) \\
$\hat \beta_1^* - \hat \beta_2^*$ & $\hat \beta_2^*$ & $\hat\beta_1^*$ & $\hat\beta_1^* -\hat \beta_2^*$ & $\hat \beta_2^*$ & $\hat\beta_1^*$ & $\hat\beta_1^* - \hat\beta_2^*$ & $\hat\beta_2^*$ & $\hat\beta_1^*$  \\ 
  \hline
0.005 & 0.002 & 0.007 & 0.071 & 0.035 & 0.072 & 0.072 & 0.036 & 0.073 & 0.140 & (0,1,2.5) & 0.265 \\ 
  0.004 & 0.005 & 0.007 & 0.072 & 0.034 & 0.073 & 0.073 & 0.034 & 0.073 & 0.140 & (0,1,2) & 0.231 \\ 
  0.004 & 0.005 & 0.007 & 0.072 & 0.034 & 0.073 & 0.073 & 0.034 & 0.073 & 0.140 & (0,0.75,1.5) & 0.231 \\ 
  0.005 & 0.002 & 0.008 & 0.125 & 0.035 & 0.125 & 0.128 & 0.036 & 0.128 & 0.050 & (0,1,2.5) & 0.150 \\ 
  0.006 & 0.005 & 0.009 & 0.125 & 0.034 & 0.125 & 0.128 & 0.034 & 0.128 & 0.050 & (0,1,2) & 0.131 \\ 
  0.006 & 0.005 & 0.009 & 0.125 & 0.034 & 0.125 & 0.128 & 0.034 & 0.128 & 0.050 & (0,0.75,1.5) & 0.131 \\ 
  0.012 & 0.002 & 0.016 & 0.181 & 0.035 & 0.180 & 0.184 & 0.036 & 0.184 & 0.025 & (0,1,2.5) & 0.105 \\ 
  0.011 & 0.005 & 0.016 & 0.184 & 0.034 & 0.183 & 0.185 & 0.034 & 0.185 & 0.025 & (0,1,2) & 0.092 \\ 
  0.011 & 0.005 & 0.016 & 0.184 & 0.034 & 0.183 & 0.185 & 0.034 & 0.185 & 0.025 & (0,0.75,1.5) & 0.092 \\ 
\hline
\hline
\end{tabular}
\caption{Mean relative bias, sample standard deviation, and mean sandwich standard error, for the estimates of $\beta_1^*$, $\beta_2^*$, and $\beta_1^* - \beta_2^*$, obtained with Method Joint-NC, over 5,000 simulated studies under the unblinded randomized controlled trial setting of Supplementary Figure \ref{webfig:DAGrandomW} and presented in Supplementary Material \ref{appendix:simulunblind}, with $n = $ 5,000. Relative bias for $\hat\beta_1^*$ is computed with respect to the log-average total effect of the vaccine on the targeted outcome. Relative bias for $\hat\beta_2^*$ is computed with respect to the log-natural indirect effect of the vaccine on the targeted outcome. Relative bias for $\hat\beta_1^* - \hat\beta_2^*$ is computed with respect to the log-natural direct effect.}\label{webtable:unblindn5000}
\end{Supplementary Table}

\end{landscape}

\newpage

\begin{landscape}

\begin{Supplementary Table}[ht]
\centering
\begin{tabular}{cccccccccccc}
\hline
\hline
\multicolumn{3}{c}{\multirow{2}{*}{Relative bias}} & \multicolumn{3}{c}{Sample standard} & \multicolumn{3}{c}{Mean sandwich}  \\
\multicolumn{3}{c}{} & \multicolumn{3}{c}{deviation} & \multicolumn{3}{c}{standard error} & P($Y_1 = 1$) & ($a_{\text{low}}, a_{\text{medium}}, a_{\text{high}}$) & corr($Y_1,Y_2$) \\
$\hat \beta_1^* - \hat \beta_2^*$ & $\hat \beta_2^*$ & $\hat\beta_1^*$ & $\hat\beta_1^* -\hat \beta_2^*$ & $\hat \beta_2^*$ & $\hat\beta_1^*$ & $\hat\beta_1^* - \hat\beta_2^*$ & $\hat\beta_2^*$ & $\hat\beta_1^*$  \\ 
  \hline
0.003 & 0.011 & $<$ 0.001 & 0.052 & 0.025 & 0.052 & 0.051 & 0.025 & 0.052 & 0.140 & (0,1,2.5) & 0.265 \\ 
  0.002 & 0.008 & 0.001 & 0.053 & 0.024 & 0.053 & 0.052 & 0.024 & 0.052 & 0.140 & (0,1,2) & 0.231 \\ 
  0.002 & 0.008 & 0.001 & 0.053 & 0.024 & 0.053 & 0.052 & 0.024 & 0.052 & 0.140 & (0,0.75,1.5) & 0.231 \\ 
  0.003 & 0.011 & $<$ 0.001 & 0.092 & 0.025 & 0.092 & 0.091 & 0.025 & 0.091 & 0.050 & (0,1,2.5) & 0.150 \\ 
  0.002 & 0.008 & $<$ 0.001 & 0.093 & 0.024 & 0.093 & 0.091 & 0.024 & 0.091 & 0.050 & (0,1,2) & 0.131 \\ 
  0.002 & 0.008 & $<$ 0.001 & 0.093 & 0.024 & 0.093 & 0.091 & 0.024 & 0.091 & 0.050 & (0,0.75,1.5) & 0.131 \\ 
  0.006 & 0.011 & 0.004 & 0.130 & 0.025 & 0.130 & 0.130 & 0.025 & 0.130 & 0.025 & (0,1,2.5) & 0.105 \\ 
  0.006 & 0.008 & 0.005 & 0.131 & 0.024 & 0.130 & 0.131 & 0.024 & 0.131 & 0.025 & (0,1,2) & 0.092 \\ 
  0.006 & 0.008 & 0.005 & 0.131 & 0.024 & 0.130 & 0.131 & 0.024 & 0.131 & 0.025 & (0,0.75,1.5) & 0.092 \\ 
\hline
\hline
\end{tabular}
\caption{Mean relative bias, sample standard deviation, and mean sandwich standard error, for the estimates of $\beta_1^*$, $\beta_2^*$, and $\beta_1^* - \beta_2^*$, obtained with Method Joint-NC, over 5,000 simulated studies under the unblinded randomized controlled trial setting of Supplementary Figure \ref{webfig:DAGrandomW} and presented in Supplementary Material \ref{appendix:simulunblind}, with $n = $ 10,000. Relative bias for $\hat\beta_1^*$ is computed with respect to the log-average total effect of the vaccine on the targeted outcome. Relative bias for $\hat\beta_2^*$ is computed with respect to the log-natural indirect effect of the vaccine on the targeted outcome. Relative bias for $\hat\beta_1^* - \hat\beta_2^*$ is computed with respect to the log-natural direct effect.}\label{webtable:unblind}
\end{Supplementary Table}

\end{landscape}

\newpage

\section{Additional remarks}\label{appendix:remarks}

\subsection{Generalized linear models}\label{appendix:models}

A few remarks on Equations (\ref{eq:equation1}) and (\ref{eq:equation2}) were already made shortly after their introduction in Section \ref{subsec:notation} in the Main Document. Three additional comments are provided below to further justify the choice of framework.

First, the conditional mean models in Equations (\ref{eq:equation1}) and (\ref{eq:equation2}) are log-linear. It is usual to consider generalized linear models with a log link function to obtain adjusted relative risks when working with a binary outcome such as $Y_1$ \cite{gail1988effect, valeri2013mediation}. Although these models can be approximated by similar models with a logistic link function when the outcomes are rare, the latter are in general employed to obtain relative odds rather than relative risks, which are precisely used to estimate vaccine efficacy \cite{orenstein1988VEdef}.

Then, while one could maybe imagine the vaccine having an interaction with a variable like age, it is much more difficult for one of the behavioral variables likely to be in $\boldsymbol{\tilde{A}}$. In the HPV vaccine literature, a reference did mention what could be seen as a potential interaction of the vaccine related to behavior variables \cite{bonneault2022vaccineinteraction}. Indeed, HPV genotype interactions depend on sexual behavior and could impact acquisition or clearance of HPV infections, and this could therefore impact the efficacy of the vaccine. However, Bonneault et al. \cite{bonneault2022vaccineinteraction} suggest that this would mostly affect non-targeted infections, that these complex mechanisms still need to be better understood, and that the potential impact would be in a longer term than that of HPV vaccine observational studies. For this reason, the model in Equation (\ref{eq:equation1}) assumes no interaction between $T$ and $\boldsymbol{\tilde{A}}$. If such an interaction were to exist and be included in Equation (\ref{eq:equation1}), the approach would no longer allow to assess the direct immunological effect of $T$ on $Y_1$, however.

Lastly, the shape of the effect of $\boldsymbol{\tilde{A}}$ on $Y_1$ and $Y_2$ is motivated by the fact that it is the more convenient for (\textit{i}) the readability of the plausible assumption of proportionality of the effect between targeted and non-targeted infections (i.e., Assumption A$_1$ given in Section \ref{subsec:nontargeted} in the Main Document); (\textit{ii}) the intuition that the probability of a woman acquiring an HPV infection is zero in the absence of sexual activity. Note that point (\textit{ii}) would be easily obtained with functions $g_1$ and $g_2$ such that $g_1(\boldsymbol{0}) = g_2(\boldsymbol{0}) = 0$, with $\boldsymbol{0}$ denoting the value of $\boldsymbol{\tilde{A}}$ in the absence of any sexual behavior.

\subsection{Choice of targeted and non-targeted infection outcomes}\label{appendix:outcomes}

In Section \ref{subsec:notation} of the Main Document, and following Etievant et al. \cite{etievantsampson2023}, $Y_1$ was defined as the binary indicator of an infection with targeted human papillomavirus (HPV) strain 16, and $Y_2 = \sum_{j=1}^{N_{NT}}Y_2^{(j)}$ as the total number of non-targeted HPV infections among the $N_{NT}$ that were recorded. Then the estimation method was based on the generalized linear models in Equations (\ref{eq:equation1}) and (\ref{eq:equation2}) in Section \ref{subsec:notation} in the Main Document or in Equations (\ref{eq:equation1w}) and (\ref{eq:equation2w}) in Section \ref{subsec:conf} in the Main Document.

As a first remark, one could focus on infections with one particular non-targeted HPV strain, but our choice of non-targeted infection outcome $Y_2$ better exploits the available data. In addition, if a model of the form in Equation (\ref{eq:equation2}) or in Equation (\ref{eq:equation2w}) holds separately for each $Y_2^{(j)}$ but with a common function $g_2$, $j \in \{1,\dots , N_{NT}\}$, then this is also true for $Y_2$. Indeed, if $$\E(Y_{2}^{(j)} \mid \boldsymbol{W}, T, \boldsymbol{\tilde{A}}) = g_2(\boldsymbol{\tilde{A}}) \exp\{\alpha_2^{(j)} + s_2^{(j)}(\boldsymbol{W})\},$$
then 
$$\E(Y_{2}\mid \boldsymbol{W}, T, \boldsymbol{\tilde{A}}) = \sum_{j=1}^{N_{NT}} \E(Y_{2}^{(j)} \mid \boldsymbol{W}, T, \boldsymbol{\tilde{A}}) =g_2(\boldsymbol{\tilde{A}})\exp\{s_2(\boldsymbol{W})\},$$
with $s_2:\boldsymbol{w}\mapsto  \log\Big[\sum_{j=1}^{N_{NT}}\exp\big\{\alpha_2^{(j)} + s_2^{(j)}(\boldsymbol{w})\big\}\Big]$. But note that it is important that the shape of the effect of $\boldsymbol{\tilde{A}}$ is the same on each $Y_{2}^{(j)}$; the magnitude of the effect can differ, however, and the proportionality constant can simply be absorbed into $\alpha_2^{(j)}$. This is a plausible assumption as cervical infections with non-targeted HPV strains share the same transmission route.

\begin{Supplementary Figure}[h!]
\centering
\includegraphics[width=0.6\linewidth]{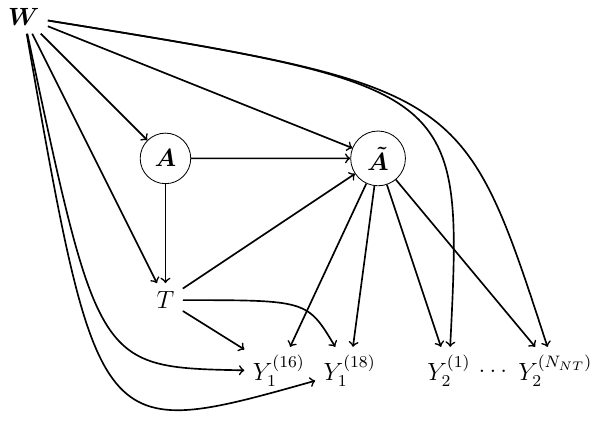}
\caption{Graph depicting the relationships among the variables in the observational setting, with measured confounders and with individual HPV infection outcomes. Circled variables are unmeasured. }\label{fig:DAGmediationconfstrains}
\end{Supplementary Figure}

Regarding the targeted infection outcome $Y_1$, one could also combine infections with several targeted HPV strains. For example, one could use $Y_1 = Y_1^{(16)} + Y_1^{(18)} - Y_1^{(16)}Y_1^{(18)}$, the binary indicator of an infection with targeted strain 16 or targeted strain 18, that Etievant et al. \cite{etievantsampson2023} named the ``composite primary outcome''. If a model of the form in Equation (\ref{eq:equation1}) or in Equation (\ref{eq:equation1w}) holds separately for $Y_1^{(16)}$ and $Y_1^{(18)}$ but with a common function $g_1$, then this is also approximately true for $Y_1$ when $Y_1^{(16)}$ and $Y_1^{(18)}$ are rare and the vaccine offers identical protection against infection by HPV 16 and infection by HPV 18. Indeed, if $$\E(Y_{1}^{(16)} \mid \boldsymbol{W},  T, \boldsymbol{\tilde{A}}) = g_1(\boldsymbol{\tilde{A}}) \exp\{\alpha_1^{(16)} + \beta_1^{(16)} T + s_1^{(16)}(\boldsymbol{W})\},$$
and
$$\E(Y_{1}^{(18)} \mid \boldsymbol{W},  T, \boldsymbol{\tilde{A}}) = g_1(\boldsymbol{\tilde{A}}) \exp\{\alpha_1^{(18)} + \beta_1^{(18)} T + s_1^{(18)}(\boldsymbol{W})\},$$
then 
\begin{eqnarray*}
\E(Y_{1} \mid \boldsymbol{W}, \boldsymbol{\tilde{A}}, T) &=& \E(Y_{1}^{(16)} \mid \boldsymbol{W}, T, \boldsymbol{\tilde{A}}) + \E(Y_{1}^{(18)} \mid \boldsymbol{W}, \boldsymbol{\tilde{A}}, T)  \\
&&- \E(Y_{1}^{(16)} \mid \boldsymbol{W}, T, \boldsymbol{\tilde{A}}) \times \E(Y_{1}^{(18)} \mid \boldsymbol{W}, \boldsymbol{\tilde{A}},  T)    
\end{eqnarray*}
as $Y_{1}^{(18)} \indep Y_{1}^{(18)} \mid \{\boldsymbol{W}, T, \boldsymbol{\tilde{A}}\}$ (see Supplementary Figure \ref{fig:DAGmediationconfstrains}). Then if $Y_1^{(16)}$ and $Y_1^{(18)}$ are rare within each stratum, $\E(Y_{1}^{(16)} \mid \boldsymbol{W}, T, \boldsymbol{\tilde{A}}) \times \E(Y_{1}^{(18)} \mid \boldsymbol{W}, T, \boldsymbol{\tilde{A}})$ is negligible compared to $\E(Y_{1}^{(16)} \mid \boldsymbol{W}, T, \boldsymbol{\tilde{A}}) + \E(Y_{1}^{(18)} \mid \boldsymbol{W}, T, \boldsymbol{\tilde{A}})$, so that $$\E(Y_{1} \mid \boldsymbol{W}, T, \boldsymbol{\tilde{A}}) \approx \E(Y_{1}^{(16)} \mid \boldsymbol{W}, T, \boldsymbol{\tilde{A}}) + \E(Y_{1}^{(18)} \mid \boldsymbol{W}, T, \boldsymbol{\tilde{A}}).$$ And if $\beta_1^{(16)} = \beta_1^{(18)} \equiv \beta_1$, $$\E(Y_{1} \mid \boldsymbol{W}, T, \boldsymbol{\tilde{A}}) \approx g_1(\boldsymbol{\tilde{A}}) \exp\{\beta_1 T + s_1(\boldsymbol{W})\},$$ with $s_1:\boldsymbol{w}\mapsto  \log\Big[\exp\big\{\alpha_1^{(16)} + s_1^{(16)}(\boldsymbol{w})\big\} + \exp\big\{\alpha_1^{(18)} + s_1^{(18)}(\boldsymbol{w})\big\}\Big]$. Again, the shape of the effect of $\boldsymbol{\tilde{A}}$ should be the same on $Y_{1}^{(16)}$ and $Y_{1}^{(18)}$ but its magnitude can differ. Note, if $Y_1^{(16)}$ and/or $Y_1^{(18)}$ are not rare, one should rather focus on an individual targeted HPV infection outcome, e.g., $Y_1 = Y_1^{(16)}$. This should also be the case if $\beta_1^{(16)} \neq \beta_1^{(18)}$, which is possible as an HPV vaccine may offer various degrees of protection to each strain.

\subsection{Heterogeneous causal effect}

Under the setting of Etievant et al. \cite{etievantsampson2023}, the causal effect of T on $Y_1$ was assumed to be homogeneous across strata of $\boldsymbol{W}$. Observe that this is not necessarily the case under the setting of Figure \ref{fig:DAGmediation} {(B)} in the Main Document. Indeed, even though at first sight in Equation (\ref{eq:equation1}) in Section \ref{subsec:notation} in the Main Document there is no interaction between $T$ and $\boldsymbol{{W}}$, an interaction could take place within $\boldsymbol{\tilde{A}}$. Notably, if one assumes $\boldsymbol{\tilde{A}} = \boldsymbol{{A}}$ in the absence of vaccination and $\boldsymbol{\tilde{A}} > \boldsymbol{{A}}$ under HPV vaccination, it is possible that the change of behavior after vaccination also depends on $\boldsymbol{{W}}$. For example, one could have $\boldsymbol{\tilde{A}} = \boldsymbol{{A}} + \delta T q(\boldsymbol{{W}})$. In other words, as protection against vaccine targeted HPV infections could be attenuated by riskier sexual behaviors following vaccination, the causal effect of the vaccine could differ across subgroups with varying degrees of risk compensation. Then, the focus could also be on the stratum-specific causal effects, to obtain a more comprehensive picture of the protection offered by the vaccine in practice.

\subsection{Causal rationale of the approach}\label{appendix:rationale}

Although the method proposed by Etievant et al. \cite{etievantsampson2023} relied on assumptions that may not hold in practice, the way infections with non-targeted strains were used was based on a strong rationale and supported by causal tools. As a reminder, they assumed log-linear models where the non-targeted strains were unaffected by HPV vaccination, and therefore knew the estimated vaccine effect on the non-targeted HPV infections should be null in the absence of confounding. Etievant et al. \cite{etievantsampson2023} thus jointly estimated the effect of vaccination on the vaccine targeted and non-targeted strains, and then ``subtracted off'' the second estimate from that on the targeted HPV strains. They showed that the confounding bias could be entirely removed under certain assumptions, notably if the unmeasured confounders affect the targeted and non-targeted strains proportionally. This is plausible as unmeasured factors creating difference between vaccinated and unvaccinated women and affecting the risk of cervical HPV infections are linked to sexual behaviors. The present work verified that when similar assumptions hold under the setting where risk compensation can take place, the quantity estimated in practice still has a clear causal meaning.

More recently, Dema et al. \cite{dema2025} also suggested using non-targeted HPV infections to control for differences in sexual behavior between vaccinated and unvaccinated women. More precisely, they included an indicator of infection with non-targeted HPV strains as a covariate in a single log-linear model estimating the vaccine effect on the targeted strains. However, Dema et al. \cite{dema2025} did not provide a rationale for their method. In particular, because the many articles discussing the use of negative control outcomes or proxies of the unmeasured confounders rely on more intricate methods (e.g., \cite{etievantsampson2023}, and \cite{shi2020NCO}), the approach proposed by Dema et al. \cite{dema2025} probably does not allow for appropriate estimation the causal effect of the HPV vaccine on the targeted strains. Future work could study the causal interpretation of the estimate of Dema et al. \cite{dema2025} under a setting such as the one in Etievant et al. \cite{etievantsampson2023} or in the current work. As a preliminary step, consider the setting without risk compensation of Etievant et al. \cite{etievantsampson2023} (i.e., when $\boldsymbol{\tilde{A}} = \boldsymbol{A}$) and when $\boldsymbol{W}=\emptyset$; see Supplementary Figure \ref{webfig:DAGbiometrics}. More precisely, assume 
\begin{eqnarray*}
    \E(Y_{1} \mid \boldsymbol{{A}}, T) &=& g_1(\boldsymbol{{A}}) \exp(\alpha_1 + \beta_1 T).
    \label{eq:equation1biometrics}
\end{eqnarray*}
As a reminder, under such a setting the average total effect of $T$ on $Y_1$ equals $\exp(\beta_1)$. But without data on $\boldsymbol{{A}}$ and in an attempt to quantify the effect of the vaccine on the targeted HPV strains, one would probably estimate $\frac{\E(Y_{1} \mid T = 1)}{\E(Y_{1} \mid T = 0)}$. Because of unmeasured confounder $\boldsymbol{{A}}$, this contrast actually equals $\exp(\beta_1) \times \frac{\E\big\{ g_1( \boldsymbol{{A}})\mid T=1\big\}}{\E \big\{ g_1(\boldsymbol{{A}})\mid T=0\big\}}$ and one would thus get a biased estimate of the causal effect. Again, under certain assumptions $\frac{\E(Y_{2} \mid T = 1)}{\E(Y_{2} \mid T = 0)} = \frac{\E\big\{ g_1( \boldsymbol{{A}})\mid T=1\big\}}{\E \big\{ g_1(\boldsymbol{{A}})\mid T=0\big\}}$ and the confounding bias can be removed by using the non-targeted infections. See Etievant et al. \cite{etievantsampson2023} for more details on the bias and on the proposed approach to reduce or remove confounding bias; also see Supplementary Material \ref{appendix:estimationinference}.

According to Dema et al. \cite{dema2025}, and instead of working jointly with $\E(Y_{1} \mid T)$ and $\E(Y_{2} \mid T)$, one could work with $\E(Y_{1} \mid T, Y_2)$. But note that 
\begin{eqnarray*}
    \E(Y_{1} \mid T, Y_2) &=& \E_{\boldsymbol{{A}}}\{\E(Y_{1} \mid T,\boldsymbol{{A}}, Y_2)\mid T, Y_2\},\\
    &=& \E_{\boldsymbol{{A}}}\{\E(Y_{1} \mid T,\boldsymbol{{A}})\mid T, Y_2\},\\
    &=& \exp(\alpha_1 + \beta_1 T) \times \E_{\boldsymbol{{A}}}\{g_1(\boldsymbol{{A}})\mid T, Y_2\},
\end{eqnarray*}
as $Y_1 \indep Y_2 \mid \boldsymbol{{A}}$; see Supplementary Figure \ref{webfig:DAGbiometrics}. Because $A \nindep T$ and $A \nindep Y_2$, the method by Dema et al. \cite{dema2025} could lead to a biased estimate of $\beta_1$. For example, even if $\E \big\{ g_1(\boldsymbol{{A}})\mid T=t, Y_2 = y_2\big\}$ can be decomposed as $\exp\{q(t) + r(y_2)\}$ for $t\in \{0,1\}$ and all possible values $y_2$ of $Y_2$, then one would estimate $\beta_1^* = \beta_1 + q(1) - q(0)$ instead of $\beta_1$.

\begin{Supplementary Figure}
\centering
\includegraphics[width=0.38\linewidth]{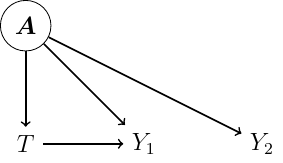}
\caption{Graph depicting the relationships among the variables in the observational setting of Etievant et al. \cite{etievantsampson2023}, without measured confounders. Circled variables are unmeasured.}\label{webfig:DAGbiometrics}
\end{Supplementary Figure}

\subsection{Alternative randomized controlled trial design to isolate immunological and behavioral effects}

With data on $\{T, Y_1\}$ from an unblinded RCT, one could estimate the total effect the vaccine on the targeted infections, but without being able to isolate the immunological and behavioral effects; see Supplementary Material \ref{sec:connection}. But if Assumption A$_1$ (given in Section \ref{subsec:nontargeted} in the Main Document) holds, using data on $Y_2$ could allow to quantify the effect of $T$ on $Y_1$ through the change of behavior only. On the other hand, estimating the behavioral effect and immunological effect from data on $\{T, Y_1\}$ only would probably require a trial with more than two arms. For example, a simple design would be a randomized trial where a third of the participants receive the vaccine and are blinded, a third of the participants receive the placebo and are blinded, and a third of the participants receive the vaccine but are not blinded. Comparing targeted infections in arms 1 and 2 would give the immunological effect of the vaccine on the targeted infections as all participants were blinded and thus would not change their behavior, while comparing arms 1 and 3 would give the behavioral effect of the vaccine as participants in arm 3 can change their sexual behavior. Such an design may be suitable for an exposure where Assumption A$_1$ is not plausible.

\newpage

\section*{Alternative text describing the figures}

\noindent Supplementary Figure \ref{webfig:conf2}: The figure contains nine panels, arranged in three columns and three rows. There is one column for the scenarios with $a_{\text{low}} = 0, a_{\text{medium}} = 0.75, a_{\text{high}} = 1.5$, one for the scenarios with $a_{\text{low}} = 0, a_{\text{medium}} = 1, a_{\text{high}} = 2$, and one for the scenarios with $a_{\text{low}} = 0, a_{\text{medium}} = 1, a_{\text{high}} = 2.5$, and there are three rows for the scenarios with $\PP(Y_1 = 1) = 0.14$, $\PP(Y_1 = 1) = 0.05$, and $\PP(Y_1 = 1) = 0.025$, respectively. The observational study size $n$ is on the $x$-axis, and the estimated log-ratio effect of the vaccine on the targeted outcome is on the $y$-axis. Each panel contains the boxplots of the 5,000 estimates with method Joint-MH in red, one for $n= 5,000$ and one for $n=10,000$, and similarly the two boxplots with method Joint-Reg in green, with method MH in cyan, and with method Reg in purple. The black horizontal solid line at $y=-0.73$ shows the true log-natural direct effect of $T$ on $Y_1$, and the black horizontal dash-dotted line at $y \approx -0.553$ or $y\approx -0.572$ shows the true log-average total effect of $T$ on $Y_1$. The log-natural direct effect of $T$ on $Y_1$ is slightly overestimated with methods Joint-MH and Joint-Reg, and heavily overestimated with methods MH and Reg.\\[-0.55cm]

\noindent Supplementary Figure \ref{fig:DAGrandom} (A): An arrow goes from $\boldsymbol{{A}}$ to $\boldsymbol{\tilde{A}}$. An arrow goes from $T$ to $Y_1$. An arrow goes from $\boldsymbol{\tilde{A}}$ to $Y_1$, and an arrow goes from $\boldsymbol{\tilde{A}}$ to $Y_2$. $\boldsymbol{{A}}$ and $\boldsymbol{\tilde{A}}$ are circled.\\[-0.55cm]

\noindent Supplementary Figure \ref{fig:DAGrandom} (B) An arrow goes from $\boldsymbol{{A}}$ to $\boldsymbol{\tilde{A}}$. A blue arrow goes from $T$ to $\boldsymbol{\tilde{A}}$, and a red arrow goes from $T$ to $Y_1$. A blue arrow goes from $\boldsymbol{\tilde{A}}$ to $Y_1$, and an arrow goes from $\boldsymbol{\tilde{A}}$ to $Y_2$. $\boldsymbol{{A}}$ and $\boldsymbol{\tilde{A}}$ are circled.\\[-0.55cm]

\noindent Supplementary Figure \ref{webfig:DAGrandomW}: Four arrows go from $\boldsymbol{{W}}$ to $\boldsymbol{{A}}$, $\boldsymbol{\tilde{A}}$, $Y_1$, and $Y_2$. An arrow goes from $\boldsymbol{{A}}$ to $\boldsymbol{\tilde{A}}$. A blue arrow goes from $T$ to $\boldsymbol{\tilde{A}}$, and a red arrow goes from $T$ to $Y_1$. A blue arrow goes from $\boldsymbol{\tilde{A}}$ to $Y_1$, and an arrow goes from $\boldsymbol{\tilde{A}}$ to $Y_2$. $\boldsymbol{{A}}$ and $\boldsymbol{\tilde{A}}$ are circled.\\[-0.55cm]

\noindent Supplementary Figure \ref{webfig:unblind}: The figure contains nine panels, arranged in three columns and three rows. There is one column for the scenarios with $a_{\text{low}} = 0, a_{\text{medium}} = 0.75, a_{\text{high}} = 1.5$, one for the scenarios with $a_{\text{low}} = 0, a_{\text{medium}} = 1, a_{\text{high}} = 2$, and one for the scenarios with $a_{\text{low}} = 0, a_{\text{medium}} = 1, a_{\text{high}} = 2.5$, and there are three rows for the scenarios with $\PP(Y_1 = 1) = 0.14$, $\PP(Y_1 = 1) = 0.05$, and $\PP(Y_1 = 1) = 0.025$, respectively. The unblinded randomized controlled trial size $n$ is on the $x$-axis, and the Joint-NC estimates are on the $y$-axis. Each panel contains the boxplots of the 5,000 $\hat \beta_1^*$ estimates in magenta, one for $n= 5,000$ and one for $n=10,000$, and similarly the two boxplots of $\hat \beta_1^* - \hat \beta_2^*$ in teal, and of $\hat \beta_2^*$ in orange. The black horizontal solid line at $y=-0.73$ shows the true log-natural direct effect of $T$ on $Y_1$, the black horizontal dash-dotted line at $y \approx -0.553$ or $y\approx -0.572$ shows the true log-average total effect of $T$ on $Y_1$, and the black horizontal dotted line at $y \approx 0.177$ or $y\approx 0.158$ shows the true log-natural indirect effect of $T$ on $Y_1$. The boxplots of $\hat \beta_1^*$ are centered around the black horizontal dash-dotted line, the boxplots of $\hat \beta_1^* - \hat \beta_2^*$ are centered around the black horizontal solid line, and the boxplots of $\beta_2^*$ are centered around the black horizontal dotted line.\\[-0.55cm]

\noindent Supplementary Figure \ref{fig:DAGmediationconfstrains} Seven arrows go from $\boldsymbol{{W}}$ to $\boldsymbol{{A}}$, $T$, $\boldsymbol{\tilde{A}}$, $Y_1^{(16)}$, $Y_1^{(18)}$, $Y_2^{(1)}$, and $Y_2^{(N_{NT})}$. An arrow goes from $\boldsymbol{{A}}$ to $T$, and an arrow goes from $\boldsymbol{{A}}$ to $\boldsymbol{\tilde{A}}$. Three arrows go from $T$ to $Y_1^{(16)}$, $Y_1^{(18)}$, and $\boldsymbol{\tilde{A}}$. Four arrow go from $\boldsymbol{\tilde{A}}$ to $Y_1^{(16)}$, $Y_1^{(18)}$, $Y_2^{(1)}$, and $Y_2^{(N_{NT})}$. Horizontal dots go from $Y_2^{(1)}$ to $Y_2^{(N_{NT})}$. $\boldsymbol{{A}}$ and $\boldsymbol{\tilde{A}}$ are circled.\\[-0.55cm]

\noindent Supplementary Figure \ref{webfig:DAGbiometrics}: Three arrows go from $\boldsymbol{{A}}$ to $T$, $Y_1$, and $Y_2$. An arrow goes from $T$ to $Y_1$. $\boldsymbol{{A}}$ is circled.

\newpage

\section*{Declarations}

\subsection*{Acknowledgments}

The author is grateful to Dean Follmann at the National Institute of Allergy and Infectious Diseases, and to the Editor and anonymous Reviewers for insightful comments on the preliminary version of this article. This work utilized the computational resources of the \href{https://gricad.univ-grenoble-alpes.fr/}{GRICAD HPC platforms}. 

\subsection*{Research ethics}

Not applicable.

\subsection*{Informed consent}

Not applicable.

\subsection*{Author contributions}

Original idea, L.E. ; Conceptualization, L.E.; Original draft preparation, L.E.; Draft review and editing, L.E. The author has accepted responsibility for the entire content of this manuscript and approved its submission.

\subsection*{Use of Large Language Models, AI and Machine Learning Tools}

None declared.

\subsection*{Conflict of Interest}

The author states no conflict of interest.

\subsection*{Research funding}

None declared.

\subsection*{Data Availability}

No data pertain to this work. R code and functions used for the simulation studies are openly available on GitHub at \href{https://github.com/Etievant/RiskCompensationNonTargetedHPV}{https://github.com/Etievant/RiskCompensationNonTargetedHPV}.

\end{document}